\pdfoutput=1
\documentclass[numberedappendix,apj]{emulateapj}
\usepackage{apjfonts}
\usepackage{xspace}
\usepackage{amsmath}
\usepackage[breaklinks,colorlinks,
  urlcolor=blue,citecolor=blue,linkcolor=blue]{hyperref}

\def\Msun{\, M_{\odot}}

\def\MH2c{\, M_{\rm H_2,cell}}
\def\Msunpc2{\,\rm M_{\odot}\,pc^{-2}}

\def\HH{{{\rm H}_2}}
\newcommand{\Rtwoh}{\mbox{$R_{200}$}}

\newcommand{\Msol}{{\,M}_\odot} 
\newcommand{\Msolyr}{{\,M}_\odot\,{\rm yr}^{-1}} 

\newcommand{\pc} {{\,\rm pc}} 
\newcommand{\K} {{\,\rm K}} 

\newcommand{\M}{{\mathcal M}}
\renewcommand{\vec}[1]{{\mathbf #1}} 
\newcommand{\kmsec}{{\,\rm {km\,s^{-1}} }}

\def\Myr{\,{\rm Myr}}
\newcommand{\kpc} {{\,\rm kpc}}

\newcommand{\fesn}{{\tt ALL\_Efb\_e001\_5ESN\ }}

\begin{document}
\shorttitle{Feedback and galaxy structure}
\slugcomment{{\em ApJ submitted}} 
\shortauthors{}

\title{The impact of stellar feedback on the structure, size and morphology\\ of galaxies in Milky Way size dark matter haloes}

\author{
Oscar Agertz\altaffilmark{1} and Andrey V. Kravtsov\altaffilmark{2,3,4}}


\altaffiltext{1}{Department of Physics, University of Surrey, Guildford, GU2 7XH, United Kingdom {\tt o.agertz@surrey.ac.uk}}
\altaffiltext{2}{Department of Astronomy \& Astrophysics, The University of Chicago, Chicago, IL 60637 USA}
\altaffiltext{3}{Kavli Institute for Cosmological Physics, The University of Chicago, Chicago, IL 60637 USA}
\altaffiltext{4}{Enrico Fermi Institute, The University of Chicago, Chicago, IL 60637 USA}

\begin{abstract}
We use cosmological zoom-in simulations of galaxy formation in a Milky Way (MW)-sized halo started from identical initial conditions to investigate the evolution of galaxy sizes, baryon fractions, morphologies and angular momenta in runs with different parameters of the star formation--feedback cycle. Our fiducial model with a high local star formation efficiency, which results in efficient feedback, produces a realistic late-type galaxy that matches the evolution of basic properties of late-type galaxies: stellar mass, disk size, morphology dominated by a kinematically cold disk, stellar and gas surface density profiles, and specific angular momentum. We argue that feedback's role in this success is twofold: (1) removal of low-angular momentum gas and (2) maintaining a low disk-to-halo mass fraction which suppresses disk instabilities that lead to angular momentum redistribution and a central concentration of baryons. However, our model with a low local star formation efficiency, but large energy input per supernova, chosen to produce a galaxy with a similar star formation history as our fiducial model, leads to a highly irregular galaxy with no kinematically cold component, overly extended stellar distribution and low angular momentum. This indicates that only when feedback is allowed to become vigorous via locally efficient star formation in dense cold gas, resulting galaxy sizes, gas/stellar surface density profiles and stellar disk angular momenta agree with observed $z=0$ galaxies.

\end{abstract} 

\setcounter{figure}{0}
\section{Introduction}
\label{sect:introduction}
The Cold Dark Matter scenario, with a low mean matter density and a cosmological constant ($\Lambda$CDM), has proven to be broadly successful in explaining and predicting a variety of observations, such as the Cosmic Microwave Background temperature anisotropies \citep[e.g][]{komatsu_etal11,hinshaw_etal13,Planckparam2013}, the evolution of cluster abundance \citep{vikhlinin_etal09b}, and the large scale distribution of matter in the Universe \citep{conroy_etal06,Springel2006}. Nevertheless, although the basic framework of galaxy formation within the CDM scenario  \citep[][]{WhiteRees78,FallEfstathiou80} is widely accepted, many aspects of the theory of galaxy formation are not yet fully understood \citep[for a recent review, see][]{Silk2012}. 

One of the most salient issues in galaxy formation modelling is understanding why galaxy formation is such an inefficient process, i.e. what controls the low fraction of cosmic baryons that end up as stars in the centers of dark matter haloes. A number of different methods, such as dark matter halo abundance matching \citep{Tasitsiomi04,vale_ostriker04,vale_ostriker06,conroy_etal06,conroy_wechsler09,Guo2010,Behroozi2013}, satellite kinematics \citep{klypin_prada09,more_etal11}, and weak lensing \citep{mandelbaum_etal06} \citep[see][for a comprehensive discussion]{kravtsov_etal14} point towards stellar to dark matter mass fractions of $M_\star/M_{\rm h}\approx 3-5\,\%$ on average for $L_\star$ galaxies and even smaller fractions for galaxies of smaller and larger mass. The fraction of baryons that ends up in observed galaxies is thus well below the cosmological baryon fraction of $\Omega_{\rm b}/\Omega_{\rm m}\approx 16\%$ \citep{Planck2015}. 

Such low baryon fractions are believed to be due to galactic winds driven by stellar feedback at the faint end of the stellar mass function \citep{DekelSilk86,Efstathiou00} and by the active galactic nuclei (AGN) and the bright end \citep{SilkRees1998,Benson2003}. Over the last two decades there has been an intense effort to incorporate and model these processes in cosmological simulations of galaxy formation  \citep[e.g.,][]{Katz92,NavarroWhite93, Katz1996, ThackerCouchman2001, Stinson06,Governato07,Scannapieco08,Colin2010,Agertz2011,AvilaReese2011,Guedes2011,Aquila,Hopkins2011,Brook2012,Agertz2013,Ceverino2013,Roskar2014,Booth2013,Aumer2013,Christensen2014,Keller2014,Kimm2015,murante_etal15}. 
However, these processes have proven to be challenging to model and results are generally mixed. In particular, until recently simulations generally produced galaxies with larger than observed baryonic masses,
and had difficulties in producing galaxies with realistic bulge-to-disk ratios \citep[although see][]{Governato2010,Guedes2011}.

Recent work \cite[e.g,][]{Stinson2013,Hopkins2014,AgertzKravtsov2015} has demonstrated the importance of stellar feedback in not only predicting the $z=0$ stellar mass-halo mass relation, but also its evolution, in particular during early stages of galaxy evolution ($z\gtrsim 2$). Furthermore, in our previous study \citep{AgertzKravtsov2015} we have shown that the interplay between star formation and stellar feedback is complex, and multiple ways exists for simulations to reproducing global galaxy observables, such as stellar masses, flat rotation curves etc. However, internal characteristics of galaxies, such as the $\Sigma_{\rm SFR}-\Sigma_{\rm gas}$ (Kennicutt-Schmidt) relation and the existence of thin galactic disks, are not reproduced by all models. 

As modeling of star formation and stellar feedback in galaxy formation simulations improves, it is important to consider, and test against, a large set of galaxy properties. In particular, it is widely recognized that in addition to basic properties of galaxies, such as stellar mass and morphology, galaxy size and angular momentum play important roles in our understanding of galaxy formation. 

In CDM models, angular momentum in galaxies is thought to originate during the initial phase of density perturbation growth, as collapsing peaks are tidally torqued by neighboring overdensities \citep{Hoyle1951,Peebles1969,Doroshkevich1970,White1984}. Angular momentum of dark matter halos is often expressed as a dimensionless spin parameter $\lambda= j/\sqrt{2}R_{\rm vir}V_{\rm vir}$ \citep[as defined by][]{Bullock2001}, which characterizes the specific angular momentum in units of the angular momentum that would be required for rotational support near the virial radius. Here  $R_{\rm vir}$ and $V_{\rm vir}$ are the virial radius and virial velocity of the halo, and $j$ the specific angular momentum inside $R_{\rm vir}$. 

In classical theories of disk galaxy formation \citep{FallEfstathiou80,ryden_gunn87,dalcanton_etal97,MoMaoWhite98}, gas acquires the same specific angular momentum as the host dark matter halo, with only a fraction of it lost during gas condensation onto the halo center. The remaining angular momentum sets the disk size, and the fraction of the angular momentum that is ultimately retained is connected to galaxy morphology and the Hubble sequence \citep[][]{Fall1983,fall_romanowsky13}. 

While the spin parameter of the dark matter  within the virial radius of halos is found to have log-normal distribution with a median $\lambda\approx 0.04$ and rms variance of $\sigma_{\ln\lambda}\approx 0.55$, regardless of halo mass and cosmic time \citep[e.g.,][]{Bullock2001,vitvitska_etal02,Bett2010}, recent studies have shown that accreted matter, dark matter as well as gas, upon entry of the virial radius, feature a higher spin than the halo's average  \citep[e.g.][]{Pichon2011,Kimm2011,Tillson2012,Danovich2012,Codis2012,Stewart2013,Ubler2014,Danovich2014}. In particular, \cite{Danovich2014} have recently used a set of zoom-in AMR simulations of galaxy formation at high redshifts ($z\gtrsim1-2$) to show that $\lambda$ of the cold gas when crossing the virial radius grows with time \citep[see also][]{Pichon2011}.

Despite the large angular momentum content of accreted gas, numerical simulations of galaxy formation have long suffered from the ``angular momentum catastrophe" \citep[][]{NavarroWhite1994,navarrosteinmetz00}, whereby galaxies forming in the centers of dark matter halos are dominated by large bulges with a significantly lower specific angular momentum content than their host haloes. 
In the last several years, galaxy formation simulations showed that this problem can be solved, or at least alleviated, by efficient stellar feedback \citep{Scannapieco08,zavala_etal08,sales_etal10}. For example, \citet[][see also \citealt{Brook2012a,Ubler2014,Christensen2014b}]{Brook2011}  found that supernova feedback can selectively remove low angular momentum gas via outflows, leading to disk formation. More recently, \cite{Genel2015} used the Illustris simulation suite to show that stellar and AGN feedback, tuned to reproduce the observed $z=0$ stellar mass function, can produce a realistic distribution of specific angular momentum of galaxies.

Despite the complexities of angular momentum evolution due to gas dynamics and effects of feedback outlined above, \cite{Kravtsov2013} showed that observed galaxy sizes and radial surface density profiles of baryons are strongly correlated with properties of their parent halo, and that sizes of both late type disks and early type spheroids appear to be set by the specific angular momentum proportional to that acquired by the dark matter halo. Subsequent studies at higher redshifts show that this finding holds for star forming galaxies from $z\sim 8$ to $z\approx 0$ \citep[e.g.,][and references therein]{Shibuya2015}. This suggests that the angular momentum distribution in galaxy disks is quite similar to the predictions of the classical disk formation models \citep[e.g.,][]{MoMaoWhite98}.

There is a number of outstanding issues related to effects of stellar feedback in setting galaxy size and angular momentum. First, while feedback can prevent the loss of angular momentum content of galaxies, or remove the lowest angular momentum gas, hence promoting the formation of disk-dominated galaxies, it is not completely clear why the baryons remaining after feedback-driven outflows   retain the angular momentum proportional to the average angular momentum of the halo. Second, as emphasized by \cite{Roskar2014}, while strong feedback may help in explaining low baryon fractions, excessively violent energy and momentum input can lead to overheating of gaseous disks, which leads to much thicker disk galaxies than are observed. 
At the very least, new feedback schemes, as they are developed and introduced, need to be tested against these basic empirical features of galaxy evolution. 

In this work  we use a suite of cosmological simulations of a Milky Way-mass halo started from the same initial conditions, but ran with different parameters for star formation and stellar feedback \citep[presented in][]{AgertzKravtsov2015}, to study the evolution of galaxy sizes, morphologies and angular momenta. Our goal is to 1) understand how different assumptions about the efficiency of star formation and stellar feedback affects the evolution of these fundamental properties of galaxies, and 2) compare these results to a variety of observed galaxy properties. 

We describe our cosmological simulations of galaxy formation in \S\,\ref{sect:method},  present our results and compare them to observational data in \S\,\ref{sect:results}, discuss our findings and compare to previous studies on this subject in \S\,\ref{sect:discussion}, and summarize our results and conclusions in \S\,\ref{sect:conclusions}.

\section{Galaxy formation simulations}
\label{sect:method}
\subsection{Star formation}
We carry out cosmological hydro+$N$-body zoom-in simulations of Milky Way mass galaxies using the Adaptive Mesh Refinement (AMR) code {\small RAMSES} \citep{teyssier02}. All simulations, and the star formation and feedback physics adopted, are presented in detail in \cite{Agertz2013} and \cite{AgertzKravtsov2015}. Briefly, we adopt a local star formation rate using the following equation:
\begin{equation}
\label{eq:schmidtH2}
\dot{\rho}_{\star}=f_{\rm H_2}\epsilon_{\rm ff}\frac{ \rho_{\rm g}}{t_{\rm ff}}, 
\end{equation}
where $f_{\rm H_2}$ is the local mass fraction of molecular hydrogen (H$_2$), $\rho_{\rm g}$ is the gas density in a cell, $t_{\rm ff}=\sqrt{3\pi/32G\rho_{\rm g}}$ the local free-fall time of the gas, and $\epsilon_{\rm ff}$ is the star formation efficiency per free-fall time. We adopt the model developed by \cite{kmt08}, \cite{kmt09}, and \cite{McKeeKrumholz2010}, hereafter the KMT09 model, for the abundance of $\HH$, based on radiative transfer calculations of idealized spherical giant atomic--molecular complexes subject to a uniform and isotropic Lyman-Werner (LW) radiation field. Relating star formation to the molecular gas is well motivated empirically, as galactic star formation rate surface densities correlate well with the surface density of molecular gas, independent of metallicity, and poorly or not at all with the surface density of atomic gas measured on kpc scales \citep{bigiel2008,Krumholz09,Gnedin09}. 

We adopt star formation efficiencies per free fall time in the range of $\epsilon_{\rm ff}\sim 0.01-0.1$, motivated by observations of local GMCs \citep{lada_etal10,Murray2011b,evans_etal14}. As demonstrated by \cite{AgertzKravtsov2015}, even large local values of $\epsilon_{\rm ff}$ ($\sim 0.1$) can reproduce the low global star formation efficiency inferred from the Kennicutt-Schmidt relation \citep[e.g.][]{bigiel2008} due to self-regulating effects of stellar feedback \citep[see also][]{Hopkins2011,Hopkins2014}.

\subsection{Stellar Feedback}
We adopt the stellar feedback model described in \cite{Agertz2013}. Briefly, each formed stellar particle is treated as a single-age stellar population with a \cite{chabrier03} initial mass function (IMF). We account for injection of energy, momentum, mass and heavy elements over time via supernova type II (SNII) and type Ia (SNIa) explosions, stellar winds and radiation pressure (allowing for both single scattering and multiple scattering events on dust) on the surrounding gas. Each mechanism depends on the stellar age, mass and gas/stellar metallicity, calibrated on the stellar evolution code {\small STARBURST99} \citep{Leitherer1999}. Feedback is done continuously at the appropriate times when each feedback process is known to operate, taking into account the lifetime of stars of different masses in a stellar population. To track the lifetimes of stars within the population we adopt the metallicity dependent age-mass relation of \cite{Raiteri1996}. 

Momentum from stellar winds, radiation pressure, and SNe blastwaves is added to the 26 nearest cells surrounding a parent cell of the stellar particle. Thermal energy from shocked SNe and stellar wind ejecta is injected directly into the parent cell. We explore the concept of retaining some fraction of the thermal feedback energy in a separate gas energy variable over longer times than expected purely form the local gas cooling time scale. This approach was discussed by \cite{Agertz2013} and \cite{Teyssier2013} \citep[for our choice of parameters, see][]{AgertzKravtsov2015}, and can be viewed as accounting for the effective pressure from a multiphase medium, where local unresolved pockets of hot gas exert work on the surrounding cold phase \citep[see recent work on superbubbles by][]{Keller2014}, or a placeholder for other sources of energy, such as turbulence and cosmic rays \citep{Booth2013}. 

Heavy elements (metals) injected by supernovae and winds are advected as a passive scalar and are incorporated self-consistently in the cooling and heating routine. The code accounts for metallicity dependent cooling by using tabulated cooling functions of \cite{sutherlanddopita93} for gas temperatures $10^4-10^{8.5}\,$K, and rates from \cite{rosenbregman95} for cooling down to lower temperatures. Heating from the UV background (UVB) radiation is accounted for by using the UVB model of \cite{haardtmadau96}, assuming a reionization redshift of  $z=8.5$. We follow \cite{Agertz09b} and adopt an initial metallicity of $Z=10^{-3}Z_\odot$ in the high-resolution zoom-in region in order to account for enrichment from unresolved Pop III star formation \citep[e.g.][]{wise_etal12}. 

\begin{table}[t]
\caption{List of cosmological zoom-in simulations. 
}
\label{table:simsummary1}

\begin{tabular}{ll}
\hline \\[-5pt]
Simulation$^a$ & Description \\ 
\\[-5pt]
\hline\\[-8pt]
\\[-8pt]
\hline
{\tt ALL\_Efb\_e001} & All feedback, $\epsilon_{\rm ff}=1\%$\\
{\tt ALL\_Efb\_e001\_5ESN} & All feedback,  $\epsilon_{\rm ff}=1\%$, $E_{\rm SNII}=5\times 10^{51}\,{\rm erg}$\\
{\tt ALL\_Efb\_e010} & All feedback, $\epsilon_{\rm ff}=10\%$
\\
\\[-8pt]
\hline
\end{tabular}
\\[2mm]
(a) All simulations adopt the KMT09 model (see main text),  and second feedback energy variable $E_{\rm fb}$ with $f_{\rm fb}=0.5$ and $t_{\rm dis}=10\Myr$. All simulations reach a minimum cell size of $\Delta x=75$~pc.
\end{table}

\subsection{Simulation suite}
\label{sect:suite}
We adopt a \emph{WMAP5} \citep{Komatsu2009} compatible $\Lambda$CDM cosmology with $\Omega_{\Lambda}=0.73$, $\Omega_{\rm m}=0.27$, $\Omega_{\rm b}=0.045$, $\sigma_8=0.8$ and $H_0=70\,{\rm km\,s}^{-1}\,{\rm Mpc}^{-1}$.  A pure dark matter simulation was performed using a simulation cube of size $L_{\rm box}=179\,{\rm Mpc}$. At $z=0$, a halo of mass $M_{\rm 200c}\approx 9.7\times10^{11}\,\Msol$ was selected for re-simulation at high resolution, and traced back to the initial redshift of $z=133$. Here $M_{\rm 200c}$ is  defined as the mass enclosed within a sphere with mean density 200 times the critical density at the redshift of analysis. The corresponding radius is $r_{\rm 200c}=205\,\kpc$. The mass within the radius enclosing overdensity  of 200 times the mean density is $M_{\rm 200m}=1.25\times10^{12}\,\Msol$ and $r_{\rm 200m}=340\kpc$. When baryons are included in the simulations, the final \emph{total} halo mass remains approximately the same.

The selected halo does not experience any major merger after $z\sim1.5$, potentially favouring the formation of an extended late-type galaxy. A nested hierarchy of initial conditions for the dark matter and baryons was generated using the {\small GRAFIC++}\footnote{{\tt http://grafic.sourceforge.net/}} code, where we allow for the high resolution particles to extend to three virial radii from the centre of the halo at $z=0$. This avoids mixing of dark matter particles with different masses in the inner parts of the domain. The dark matter particle mass in the high resolution region is $m_{\rm DM}=3.2\times 10^5\,\Msol$ and the adaptive mesh is allowed to refine if a cell contains more than eight dark matter particles, and similar criterion is employed for the baryonic component. At the maximum level of refinement, the simulations reach a physical resolution of $\Delta x\approx75\,\pc$.

In this work we focus on three zoom-in simulations from the suite presented in \citet{AgertzKravtsov2015}.  All models adopt the same feedback and star formation implementation, but with diffferent parameters, and are picked to represent three different galaxy formation scenarios: 1) overcooling due to inefficient feedback ({\tt ALL\_Efb\_e001}), 2) efficient feedback via boosted supernova feedback ({\tt ALL\_Efb\_e001\_5ESN}), and 3) efficient feedback via efficient star formation ({\tt ALL\_Efb\_e010}). 

In {\tt ALL\_Efb\_e001} the efficiency of star formation is $\epsilon_{\rm ff}=1\%$. \citet{AgertzKravtsov2015} demonstrated how this resulted in inefficient feedback incapable of driving global galactic outflows. This inability results leads to overproduction of stars at high redshifts, in strong tension with semi-empirically derived star formation histories \citep[e.g.][]{Moster2013,Behroozi2013}. This run forms a very dense and rapidly rotating central concentration (see Section \ref{sect:results_am} and Figure \ref{fig:vcirc} below), which makes the simulation computationally extensive. Given that the galaxy it produces was clearly unrealistic, this simulation was run only to $z=1.5$. 

For {\tt ALL\_Efb\_e001\_5ESN} the overcooling problem was resolved by boosting the released energy per supernova by a factor of five. This led to a good match to many considered observables at $z\gtrsim1-2$: semi-empirically derived star formation histories, the stellar mass-gas metallicity relation and evolution, the stellar mass-halo mass ($M_\star-M_{\rm 200}$) relation and its evolution, as well as flat shapes of rotation curves. However, the normalization of the Kennicutt-Schmidt relation was in tension with high redshift data, and the vigorous galactic outflows ended up preventing the formation of a cold galactic disk at any redshift, as we show below.

{\tt ALL\_Efb\_e010} explores a scenario where vigorous galactic outflows are generated not by artificially boosting the available stellar feedback momentum or energy budget, but by adopting a higher local star formation efficiency per free-fall time, here $\epsilon_{\rm ff}=10\%$, in agreement with observed values in massive GMCs \citep[e.g.][]{lada_etal10,Murray2011b,evans_etal14}, which leads to more correlated stellar feedback events. This resulted in galactic properties in good match with all considered observables at $z\gtrsim1$, while producing galactic wind mass loading factors that decreases with increasing dark matter halo masses, leading to an epoch of disk formation at $z\lesssim1$. In this work we continue the analysis down to $z=0$.
 
The simulations used in this study and their star formation and feedback parameters, are summarized in Table \ref{table:simsummary1}.

\begin{figure}[t]
\begin{center}
\includegraphics[width=0.48\textwidth]{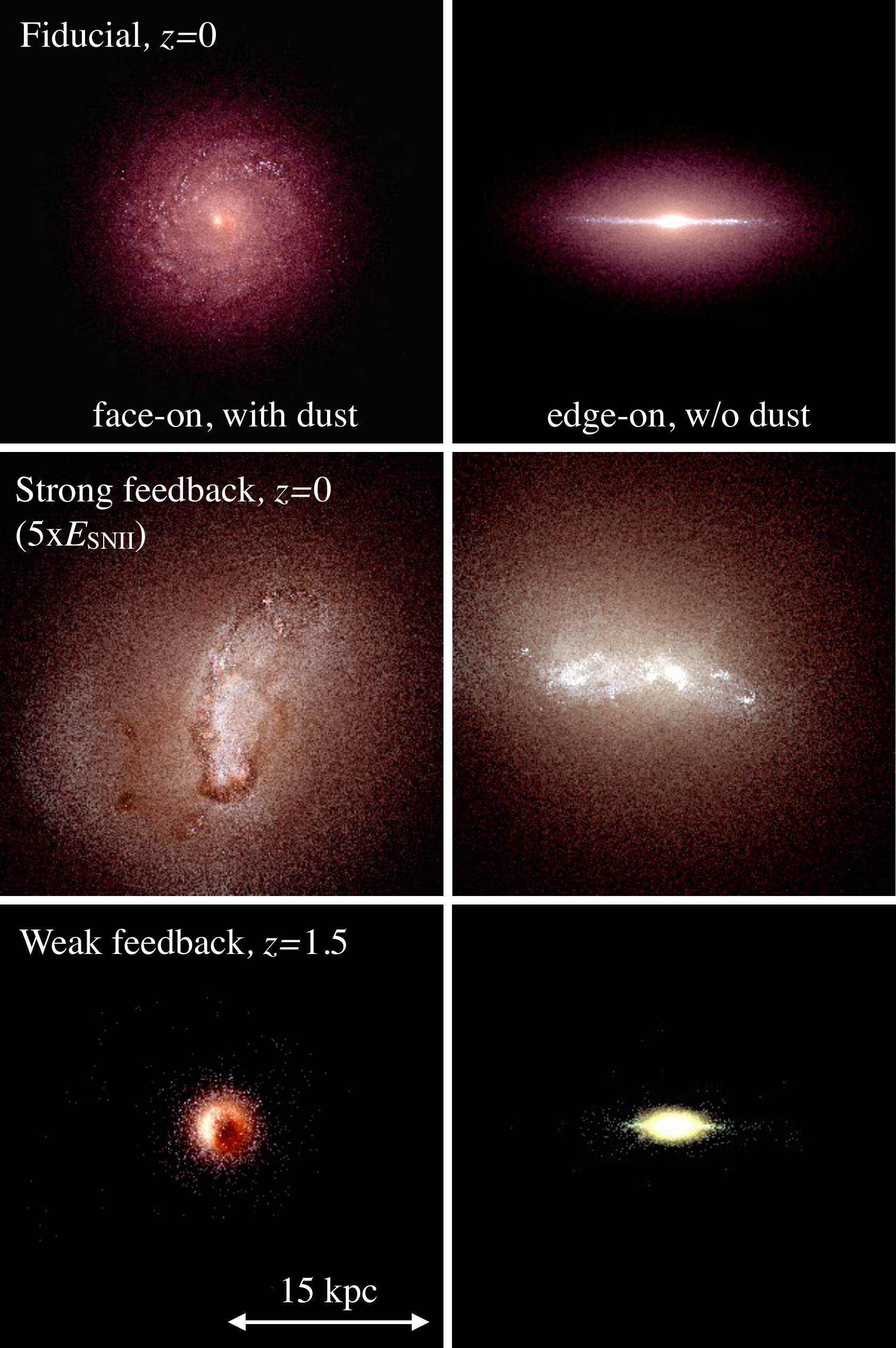}
\caption{Mock $g,r,i$ composite images of the three simulations at their final output, shown face-on in the left column and edge-on in the right column. The images are constructed assuming single stellar population SEDs from STARBURST99 \citep{Leitherer1999} for each star particle, given its known age and metallicity, and then ray-tracing the line-of-sight flux. In the face-on images, the flux is attenuated using a MW-like reddening curve \citep{WeingartnerDraine2001} with a dust-to-gas ratio vs. gas metallicity relation compatible with \cite{Fisher2014}. The galaxy in the run with low star formation efficiency (bottom row) features a very compact morphology at $z=1.5$, in stark contrast to the fiducial model which features a well defined galactic disks at $z=0$ (upper row). The model with boosted energy per supernovae (middle row) fails to form a thin disk, as star formation occurs in an extended and highly turbulent gaseous system. 
}
\label{fig:disks}
\end{center}
\end{figure}

\begin{figure}
\begin{center}
\includegraphics[width=0.45\textwidth]{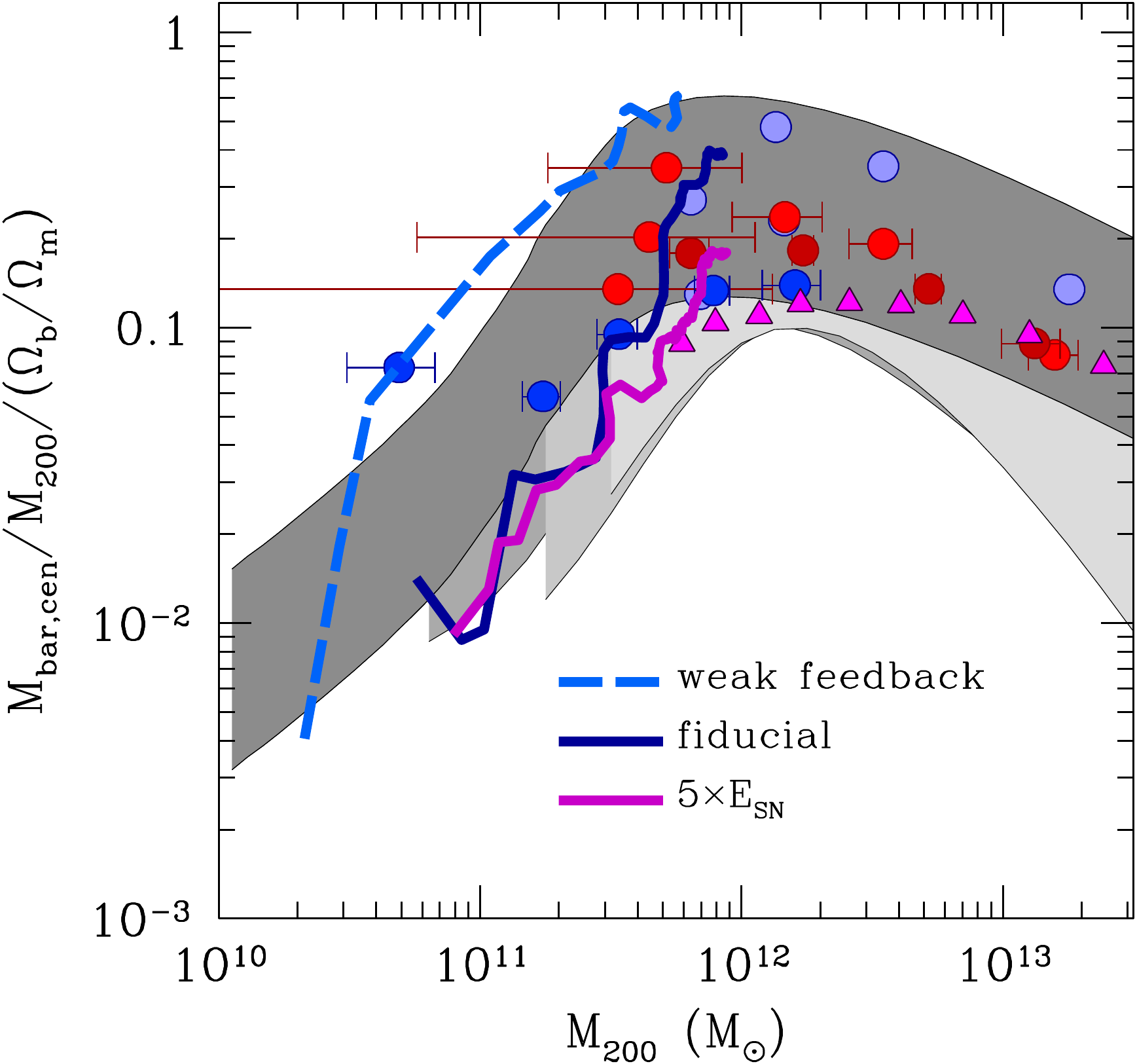}\quad\quad
\caption{The evolution of the stellar mass fraction in units of the universal baryon fraction as a function of halo mass. The shaded regions show, from light to dark grey, the $z=3,2, 1$ data from \cite{Behroozi2013} and $z=0$ from \cite{kravtsov_etal14}, where the gray bands show the region $\pm 2\sigma$ around the mean relation. The points show observational inferences of stellar fractions for galaxies at $z\approx 0$ using weak lensing (circles) for late type (blue) and early type (red) systems by \citet{mandelbaum_etal06} and \citet{Hudson2015}. Magenta triangles show estimates using satellite kinematics by \citet{more_etal11}.}
\label{fig:lfms}
\end{center}
\end{figure}

\section{Results}
\label{sect:results}
\subsection{Morphology}
\label{sect:Morphology}

Figure \ref{fig:disks} shows mock composite $g,r,i$-filter face-on and edge-on images of galaxies in the three simulations at their respective final redshifts, which account for dust attenuation.  
As we noted in the previous section, the simulation with low star formation efficiency ({\tt ALL\_Efb\_e001}, bottom row) was not evolved to $z=0$ due to the high computational expense, as the galaxy suffered from the classical overcooling problem (see \S\,\ref{sect:results_baryon}) and was stopped at $z=1.5$.
While the acquisition of galactic angular momentum, and diversity of galaxy morphologies, ultimately is linked to properties of cosmological flows and the halo merger history (see discussion in \S\,\ref{sect:introduction}), we here find that despite \emph{identical} initial conditions, the resulting galactic morphologies are strikingly different due to relatively small changes in the parameters of small-scale star formation and feedback \footnote{We note that we are only concerned with galaxies forming in $\sim 10^{12}\Msun$ haloes, and not massive galaxies in the high-$z$ Universe which likely are progenitors of today's early type galaxy population \citep[][]{Zolotov2015}. In massive disks at high $z$, properties of cold stream accretion, such as accretion alignment, likely matter more than for the galaxy types investigated here \citep[e.g.][]{Danovich2014}}.

Overcooling in the {\tt ALL\_Efb\_e001} run (bottom row) produces a very massive galaxy and compact morphology with no clear signature of an extended stellar disk at any time. This morphology is in stark contrast with the well defined galactic disk formed in the fiducial model ({\tt ALL\_Efb\_e010}, top row), which features both a thick and thin stellar disk component at $z=0$. The vertical velocity dispersion of the young ($t_\star<3$ Gyr) thin disk is $\sigma_{\star,z}\sim 10\kmsec$ \citep[for a full analysis, see][]{AgertzKravtsov2015}, and $\sim60-70\kmsec$ for the old thick disk, compatible with what is observed in the Milky Way \citep[e.g.][]{Bovy2012}. The model with low star formation efficiency, but boosted energy release per supernova ({\tt ALL\_Efb\_e001\_5ESN}), fails to form a cold thin disk, with star formation occuring in an extended and highly turbulent gaseous system. The overall morphology is irregular, with large gaseous clumps and irregular dark dust lanes.

\subsection{Baryon fractions}
\label{sect:results_baryon}
Figure \ref{fig:lfms} shows the stellar mass fraction of the central galaxies in the three runs, expressed in units of the cosmic baryon fraction, as a function of dark matter halo virial mass for the three models. We compare the simulation data to results from semi-empirical constraints based on the abundance matching approach \citep{Behroozi2013,kravtsov_etal14}, as well as inferences from weak lensing studies of late and early type systems \citep{mandelbaum_etal06,Hudson2015} and estimates using satellite kinematics \citep{more_etal11}.
 
The fiducial and the boosted feedback runs conform to the $z=0$ stellar mass -- halo mass relation, and its inferred evolution. As already discussed in \cite{AgertzKravtsov2015}, these models achieve this in very different ways leading to dramatic differences in the resulting galaxy morphologies (Figure \ref{fig:disks}). The run with 
low local star formation efficiency is inconsistent with high redshift data, and is marginally consistent with the expected scatter of stellar fractions at $z=0$, although note that the final data point is at $z=1.5$.

\subsection{Size evolution in simulations and comparison with observations}
\label{sect:results_size}
\begin{figure}[t!]
\begin{center}
\includegraphics[width=0.45\textwidth]{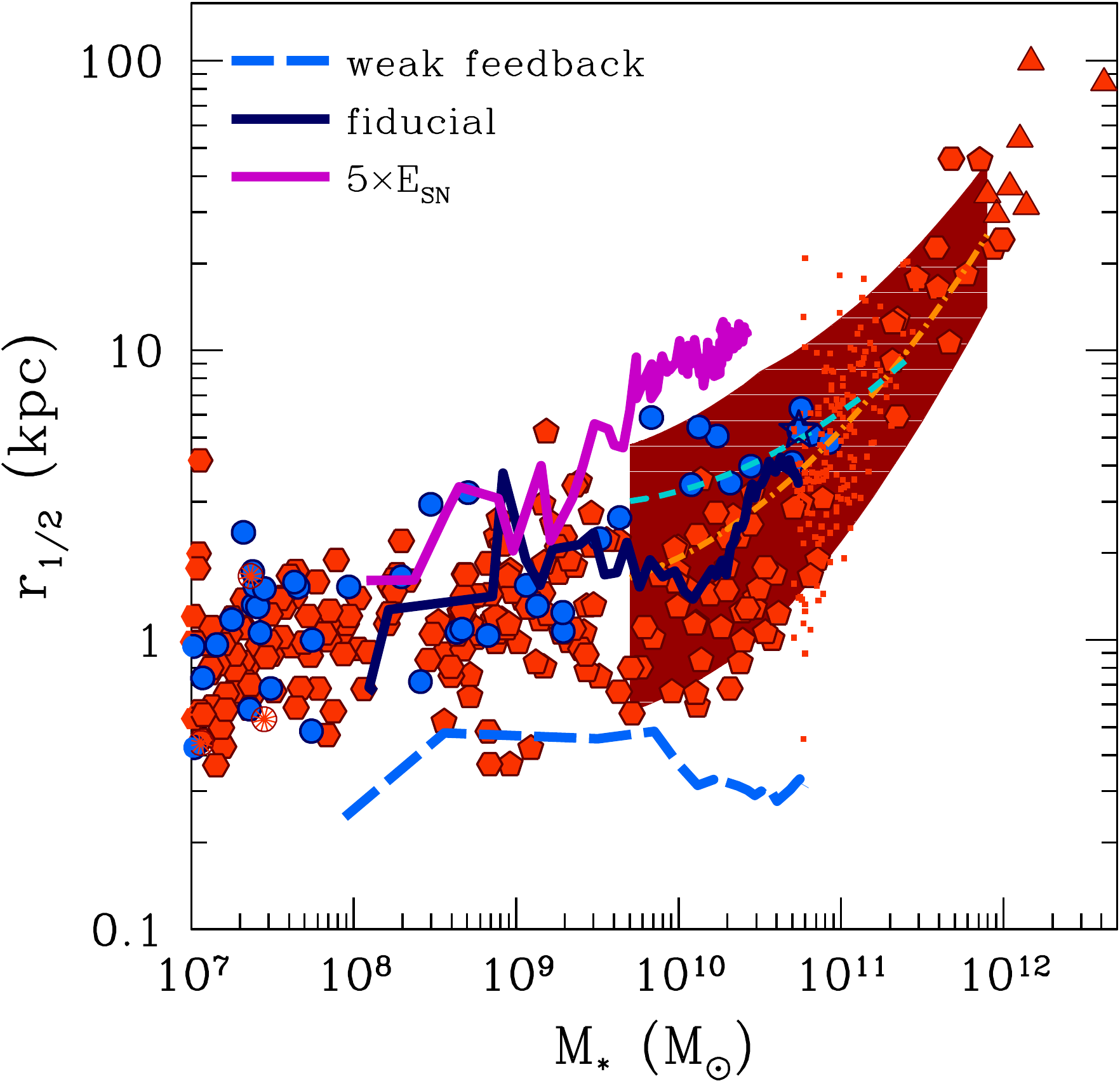}
\caption{Evolution of the stellar 3D mass-half mass radius relation for galaxies in our three runs (lines) against the observed relation at $z=0$ (points) for galaxies of different morphological types.   The simulations are shown by the solid blue (fiducial), magenta (\fesn), and dashed light blue line (low star formation efficiency, weak feedback) lines. Observed galaxies shown by points of the following type: the {\it red pentagons} and {\it hexagons} show a sample of elliptical and dwarf elliptical galaxies from the compilation of \protect\citet{misgeld_hilker11}; {\it blue circles} are the late type galaxies from the samples of \protect\citet{leroy_etal08} and \citet{zhang_etal12} with half-mass radii estimated from the deprojected surface density profiles, while the {\it star symbol} shows the Milky Way; the {\it red cartwheel} points show the Local Group dwarf spheroidal galaxies from the compilation of \protect\citet{misgeld_hilker11}. The {\it light blue dashed line} and {\it dot-dashed orange line} show the average relations derived by \protect\citet{bernardi_etal12} for late and early-type galaxies, respectively. {\it Dark red shaded band} shows $2\sigma$ scatter around the mean relation calculated for all galaxies in the \citet{bernardi_etal12} sample. Individual galaxies from the SDSS, presented in \protect\citet{szomoru_etal12}, are shown by dots.}
\label{fig:rhms}
\end{center}
\end{figure}

\begin{figure*}[t]
\begin{center}
\begin{tabular}{cc}
\includegraphics[width=0.45\textwidth]{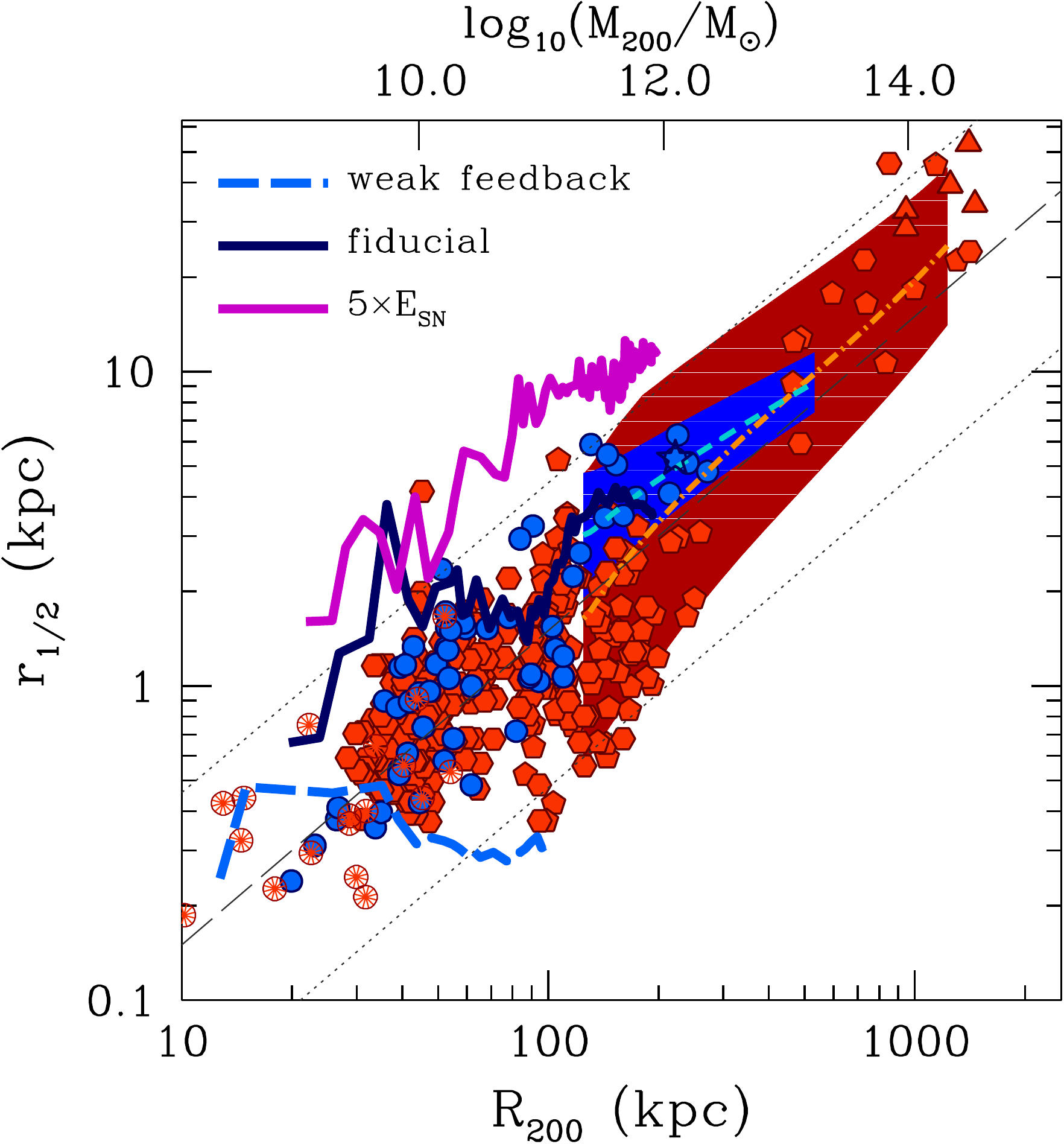}
\includegraphics[width=0.45\textwidth]{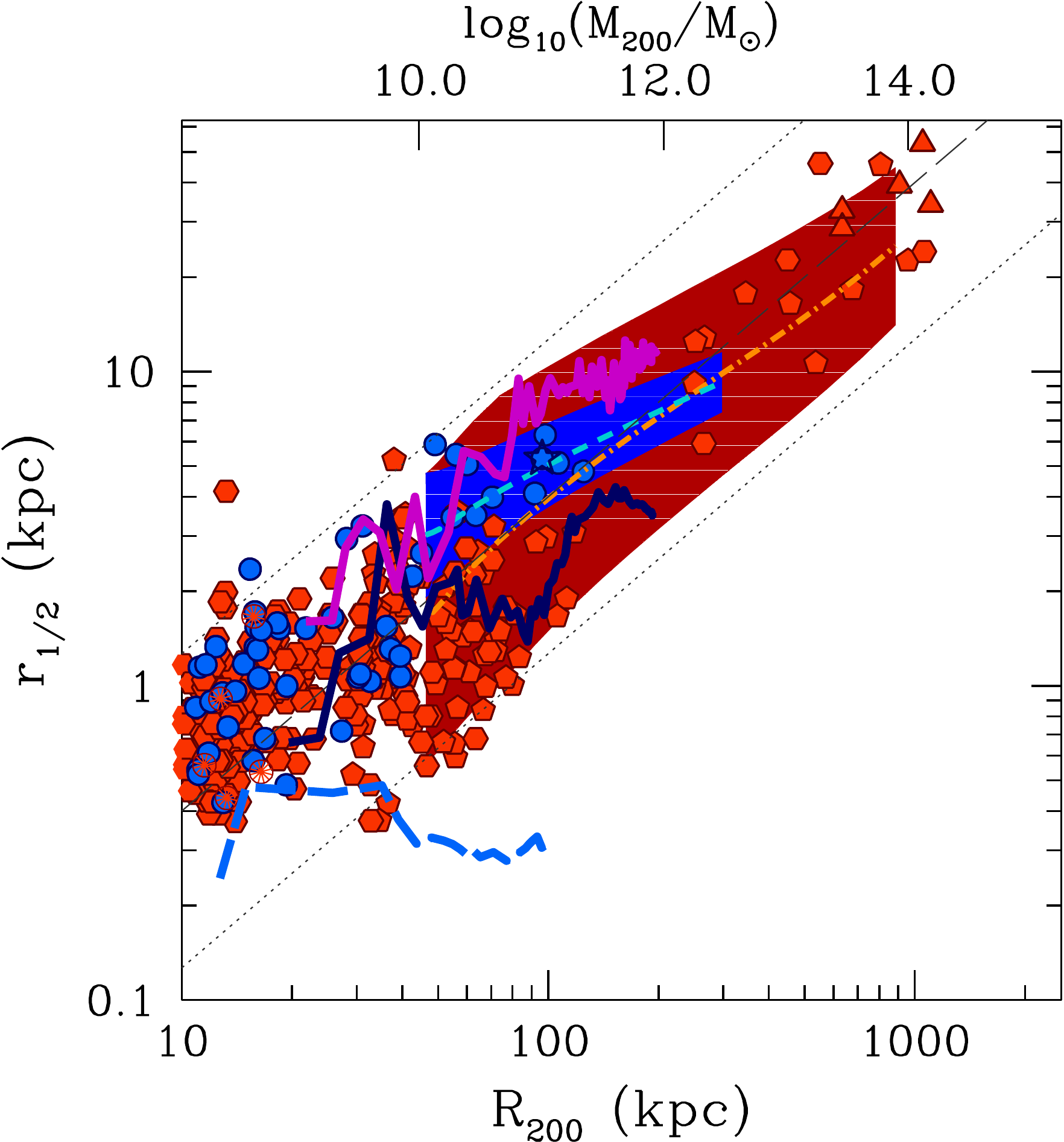}
\end{tabular}
\caption{Left: 3D half-mass radius-- $R_{200}$  radius relation of galaxies, where $R_{200}$ is the radius enclosing density contrast of 200 relative to the critical density. Right: the same relation but with $R_{200}$ of observed galaxies corrected approximately for pseudo-evolution (see text for details). The simulations are shown by the solid blue (fiducial), magenta (\fesn), and dashed light blue line (low star formation efficiency, weak feedback) lines, which are identical in both panels. 
The $R_{200}(z)$ radius is measured for the simulated galaxies at the redshift of analysis. As shown by \protect\citet{Kravtsov2013}, the observed galaxies at $z\approx 0$ on average follow a nearly linear relation, $r_{1/2}\approx 0.015R_{200}$ 
spanning more than eight orders of magnitude in stellar mass. The {\it red pentagons} and {\it hexagons} show a sample of elliptical and dwarf elliptical galaxies from the compilation of \protect\citet{misgeld_hilker11}; {\it blue circles} are the late type galaxies from the samples of \protect\citet{leroy_etal08} and \citet{zhang_etal12} with half-mass radii calculated directly from the deprojected surface density profiles, while the {\it star symbol} shows the Milky Way; the {\it red cartwheel} points show the Local Group dwarf spheroidal galaxies from the compilation of \protect\citet{misgeld_hilker11}. The {\it light blue dashed line} and {\it dot-dashed orange line} show the average relations derived for late and early-type galaxies, respectively, from the average $R_{\rm 1/2}-M_\star$ relations of \protect\citet{bernardi_etal12}. {\it Dark red shaded band} shows $2\sigma$ scatter around the mean relation calculated for all galaxies in the \citet{bernardi_etal12} sample. {\it The orange dot-dashed line} with error bars shows the mean relation and $2\sigma$ scatter for massive SDSS galaxies presented in \protect\citet{szomoru_etal12}; individual galaxies from this sample are shown by blue (S\'ersic index $n<2.5$) and red ($n>2.5$) dots. {\it The gray dashed line} shows linear relation $r_{1/2}=0.015\Rtwoh$ and dotted lines are linear relations offset by 0.5 dex, which approximately corresponds to the scatter in galaxy sizes from distribution of  halo spin parameter $\lambda$ under assumption that $r_{1/2}\propto\lambda\Rtwoh$.}
\label{fig:rgrh}
\end{center}
\end{figure*}

Figure~\ref{fig:rhms} shows evolution of the stellar mass-half mass radius ($M_\star-r_{1/2}$) relation for galaxies in our three runs against the observed relation at $z=0$ \citep[see][for details about observational sample and the method used to estimate $r_{1/2}$]{Kravtsov2013}. The low efficiency star formation run evolves off the mean relation at early times, and at the last available data point ($z=1.5$) the galaxy is extremely compact with $r_{1/2}\lesssim 0.5\kpc$. We note that a significant fraction of $z\sim1-3$ galaxies are in fact compact ellipticals \citep{vandokkunm2008,Newman2010,vandokkunm2014} with $r_{1/2}< 1\kpc$ \citep{Carollo2013}. However, such galaxies are believed to be progenitors of today's elliptical galaxies that 
evolve at low $z$ primarily due to mass accretion via minor mergers \citep[e.g.,][]{Hopkins2009}. The compact ellipticals are mostly quenched at low $z$, as characterized by their specific star formation rates. With a SFR$\sim 20\Msolyr$ at $z=1.5$, the overcooling model is far from being quenched, and while it could be identified as a ``blue nugget" \citep[progenitors of compact ellipticals, see e.g.][]{Barro2013,DekelBurkert2014}, and join the population of early type galaxies at $z=0$ by quenching and late time expansion \citep[][]{Whitaker2012}, its already significant stellar mass fraction makes it a rather unlikely candidate for a realistic galaxy. 

In contrast, the boosted feedback run {\tt ALL\_Efb\_e001\_5ESN}, in which  the resulting galaxy lacks a cold disk component, is very extended, with $r_{1/2}>10\kpc$ at $z=0$. This is $\sim 2-3$ times greater than the observed average $r_{1/2}$. Interestingly, extended spheroidal systems with exponential surface brightness profiles do exist in the nearby universe \citep{vandokkum_etal15}. However, such galaxies are rather rare and
cannot be considered to be the typical outcome of galaxy formation in a $\approx 10^{12}\ \rm M_{\odot}$ halo. 

Figure~\ref{fig:rhms} also shows that the galaxy forming in the fiducial run matches the $z=0$ observations at all times. Prior to the last major merger at $z\sim 1.8$, when $M_\star\approx 10^{10}\,\Msun$, $r_{1/2}$ is roughly constant at $\sim 1-2\,\kpc$. After this point the galaxy enters an epoch of disk formation and grows rapidly in size, with $r_{1/2}\sim 3-4\,\kpc$ at $z=0$, close to the estimated value for the Milky Way \citep{BovyRix2013}. More precisely, starting from $M_\star\approx10^{10}\Msol$ at $z\sim1.8$, our fiducial model grows by a factor of $\sim 1.75$ in $r_{1/2}$ while growing by a factor of 6 in mass until $z=0$, indicating $r_{1/2}\propto M_\star^{0.3}$. This is in excellent agreement with the the results of \cite{vandokkum2013} and \cite{Patel2013}, who studied the structural evolution of MW-mass progenitors back to $z\sim 2.5$ using the 3D-HST and CANDELS Treasury surveys. Here the half-light radii of star forming galaxies in their sample were found to scales as $r_{1/2}\propto M_\star^{0.29\pm0.08}$.

These results illustrate that while the galactic environment, via mergers, alignment of accretion etc., must play a role for the resulting galaxy size and morphology \citep[especially in explaining the origin of early type galaxies, e.g.,][]{Hopkins2010}, internal small-scale feedback processes, and our limited theoretical understanding of how these should be accounted for over cosmic time, impacts, and possibly even dictates, the Hubble sequence of late type galaxies. We discuss this further in \S\,\ref{sect:discussion}.

\begin{figure*}[t]
\begin{center}
\begin{tabular}{ccc}
\includegraphics[width=0.5\textwidth]{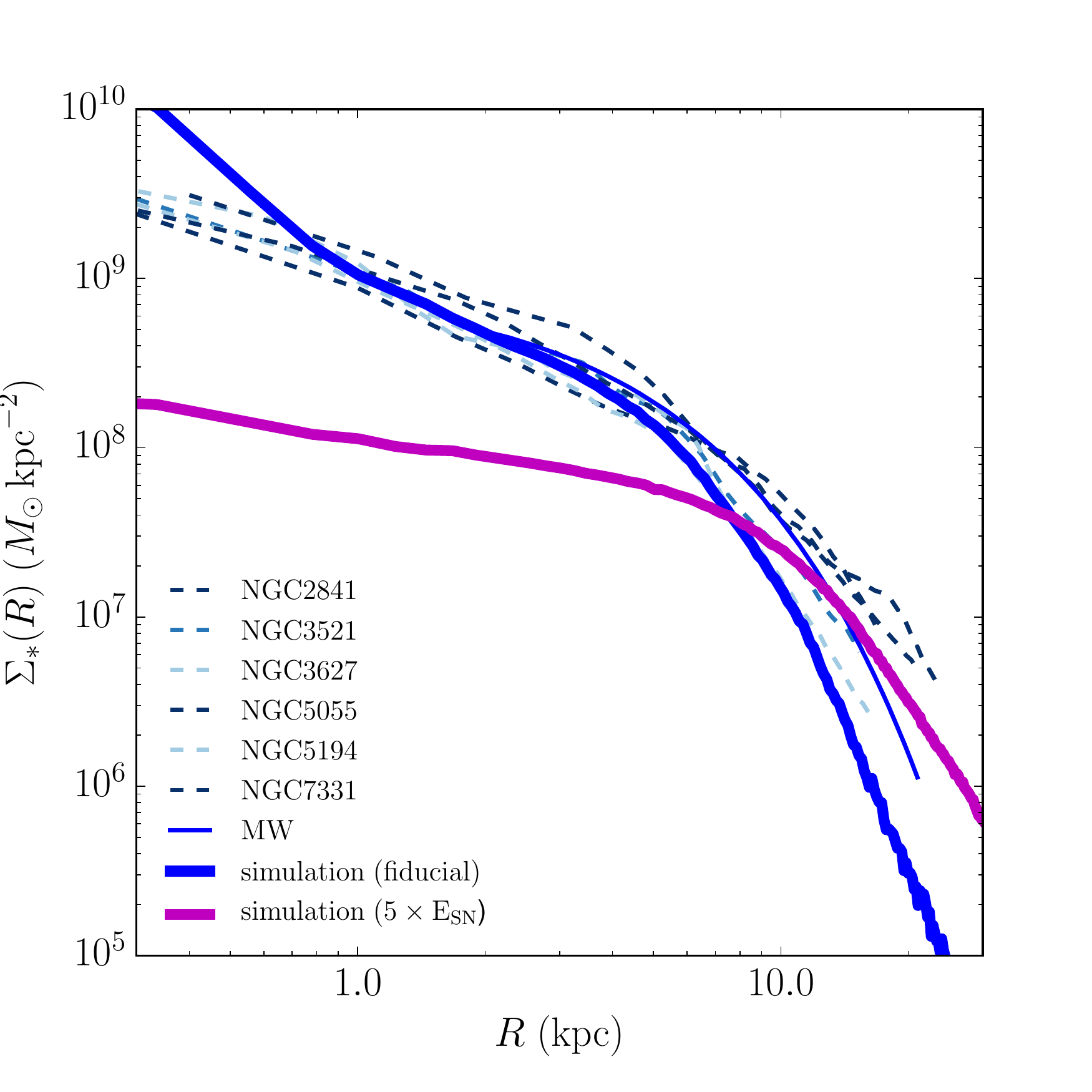}
\includegraphics[width=0.5\textwidth]{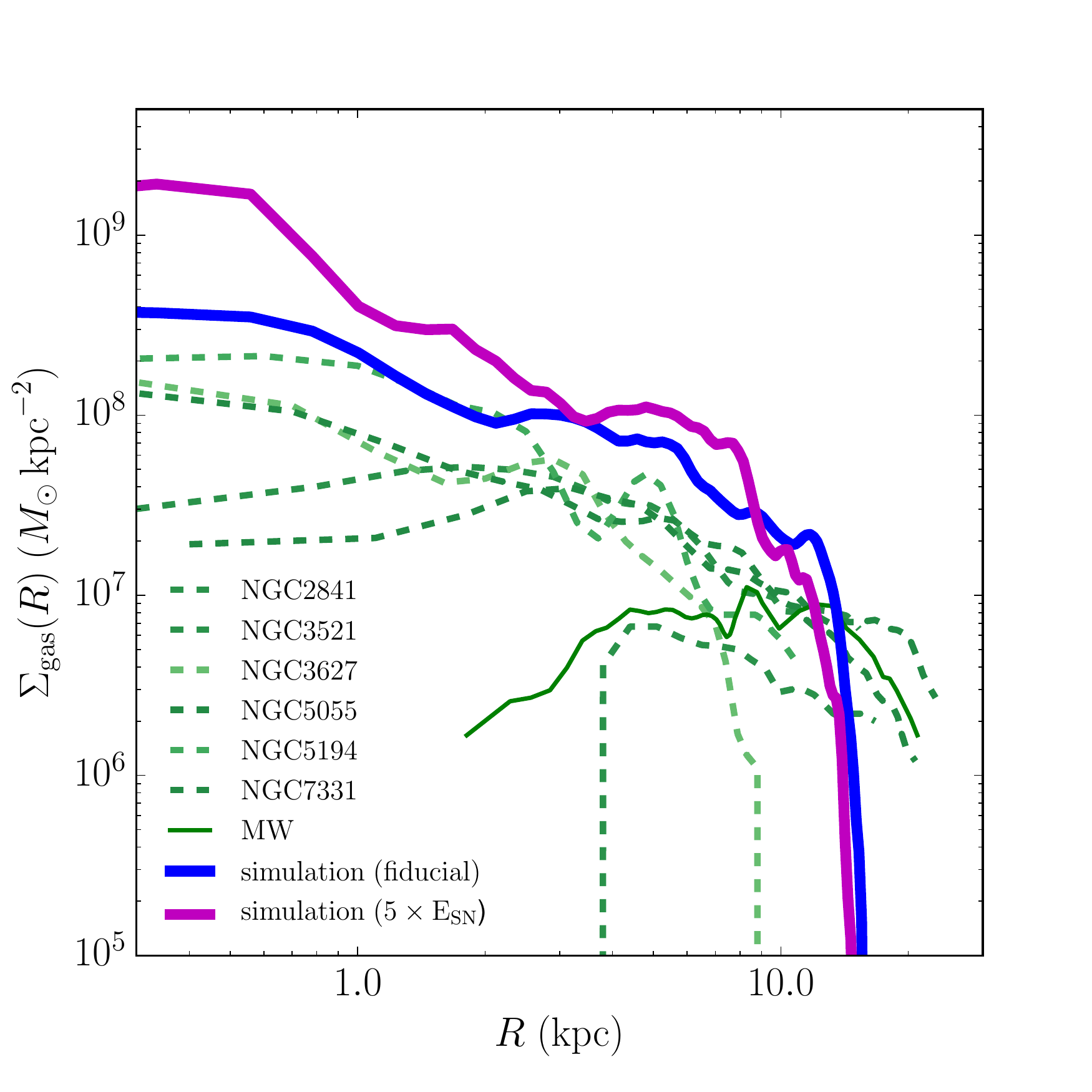}
\end{tabular}
\caption{Comparison of the surface density profiles of stars (left) and cold gas (right) in the simulated galaxies in the fiducial and \fesn simulations along with the profiles of late type galaxies from the THINGS sample \protect\citep{leroy_etal08} in the stellar mass range $M_\star\approx 3-10\times 10^{10}\ \rm M_{\odot}$ and the Milky Way, using  a combination
of the thin and thick stellar disks with parameters given in
Table 2 of \protect\citet{McMillan2011} for $\Sigma_*$ and the sum of HI and H$_2$ profiles from \protect\citet{Dame1993} for $\Sigma_{\rm gas}$.  In observed galaxies the gas surface density includes atomic and molecular gas corrected for Helium mass fraction, while in simulations the gas mass includes all of the cold gas of $T<10^4$ K.}
\label{fig:ssgpro}
\end{center}
\end{figure*}

The left panel of Figure \ref{fig:rgrh} shows the evolution of galaxies in the three runs in the plane of half-mass radius, $r_{1/2}$, vs. the radius, $R_{200}$, enclosing the density contrast of 200 relative to the critical density of the universe. Simulation results are shown against a collection of $z\approx 0$ observed galaxies (see caption for the data sources), for which $R_{200}$ was estimated from the halo mass obtained by abundance matching. As shown by \citet{Kravtsov2013}, this relation is on average close to linear, $r_{1/2}\approx 0.015\,R_{200}$, which is shown by the long gray dashed line along with the $2\sigma$ scatter, as expected if galaxy sizes were set by the specific angular momentum of baryons shared with the dark matter halo (i.e., $r_{1/2}\propto \lambda R_{200}$). A similar relation has since been observed at a wide range of redshifts \citep[e.g.,][]{Shibuya2015}. 

The figure shows that in the weak feedback simulation the galaxy maintains a compact distribution of stars with half-mass radius staying constant at $r_{1/2}\sim 0.3-0.5$ kpc. The fiducial and \fesn simulations, however, evolve approximately along the linear relation, albeit with significant fluctuations, especially during early stages evolution (smaller sizes). As discussed above, the galaxy in the \fesn run has a size that is too large for a typical halo expected to host its galaxy at $z=0$. Such galaxies are quite rare in the observed samples. The fiducial run is close to the $z=0$ data inferred for observed galaxies during the late stages of evolution. At high $z$, the ratio $r_{1/2}/R_{200}$ is about twice larger than at low $z$, a behavior that is consistent with the average behavior of observed galaxies reported by  \citet[][]{Shibuya2015}. Such evolution may be partly due to the physical evolution of the disk size, but could also arise because
the disk size is set during the fast mass growth regime of halo evolution after which a large fraction of the $M_{200}$ growth is not a real physical growth but an increase of mass due to pseudo-evolution \citep{Diemer2013}, and is hence not accompanied by the proportional growth of stellar $r_{1/2}$. 

To illustrate the latter point, the right panel of  Figure \ref{fig:rgrh} shows a somewhat different representation of the evolution. Here, the curves representing tracks of simulated galaxies are unchanged from the left panel. The observed $z=0$ sample, however, is corrected for possible {\it pseudo-evolution} (PE) of the halo mass. Many indications exist that the inner regions of halos change little from $z\sim 1-2$ until the present \citep{cuesta_etal08,Diemer2013,More2015}. Thus, the evolution of $M_{200}$ and $R_{200}$ at these redshifts may be simply due to the changing critical density of the universe, to which they are tied by definition, and not by the real physical growth. In the extreme scenario, one can imagine that halo grows only by pseudo-evolution since its formation time, which we can operationally define as the time when the halo had a concentration of $c_{\rm form}=3$ \citep{Wechsler2002,Zhao2009}.
Thus, in the right panel we correct the virial radii as $R_{200}^\prime=c_{\rm form}R_{200} /\bar{c}(M_{200},z=0)$, where $\bar{c}(M_{200},z=0)$ is the average concentration of halos of mass $M_{200}$ at $z=0$, which we calculated using an accurate concentration model by \cite{DiemerKravtsov2015} calibrated on a large suite of simulations. 

If galaxy halos do exhibit significant pseudo-evolution, the disk sizes were set at higher $z$ and thus they should be compared to the PE-corrected $R_{200}$ values. In other words, in such case it should be more appropriate to compare higher-$z$ portion of the galaxy tracks with the data
with the PE-corrected virial radii. The right panel of Figure \ref{fig:rgrh} indeed shows that the early stages of galaxy evolution, and hence lower progenitor stellar masses and halo virial radii, track the PE-corrected data points. At late stages they evolve close to the $z\approx 0$ data. This evolution qualitatively and quantitatively tracks the evolution derived for observed galaxies across a wide range of redshifts by \citet{Shibuya2015}, who find that star forming galaxies have $r_{1/2}/R_{200}\approx 0.015-0.02$ at $z\lesssim 2$   increasing somewhat to $r_{1/2}/R_{200}\approx 0.03-0.04$  at larger redshifts. 

Overall, these results indicate that in the fiducial and \fesn simulations, the angular momentum of the disk is approximately linearly proportional to the specific angular momentum of the halo and galaxies thus evolve along the $r_{1/2}\propto R_{200}$ track. Similar results was recently reported by \cite{PedrosaTissera2015}, who found that galaxies in a large simulated sample followed a linear $R_{200}-r_{1/2}$ relation
in simulation with efficient feedback, but disk and spheroidal galaxies followed relations with different constants of proportionality.

\begin{figure*}[t]
\begin{center}
\includegraphics[width=\textwidth]{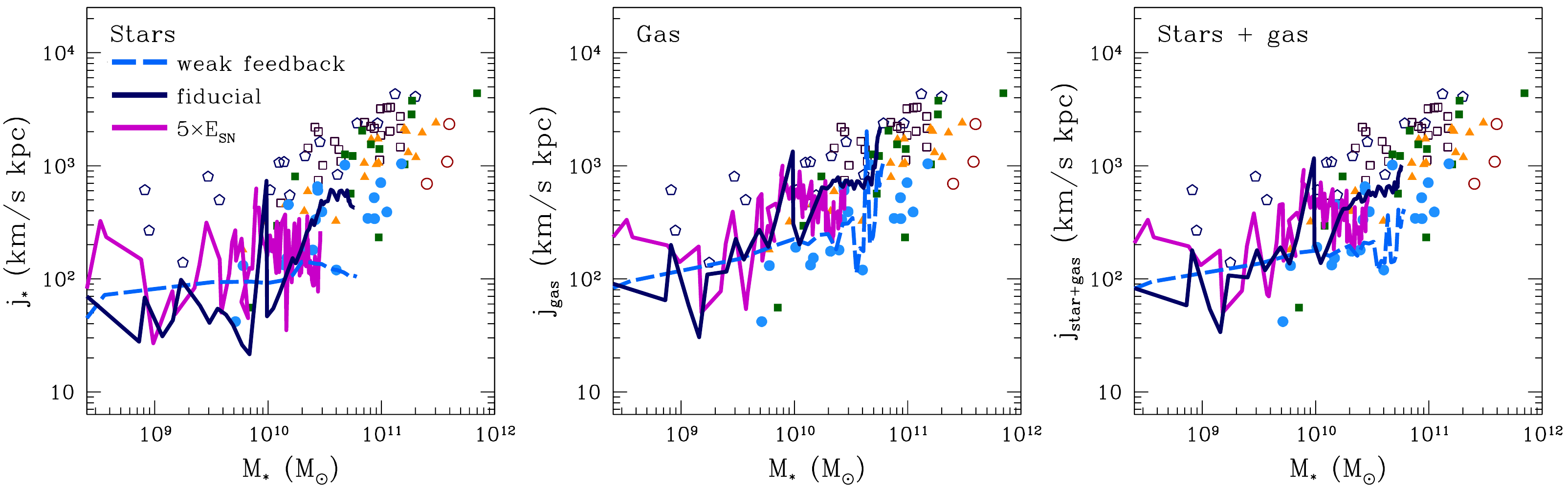}
\caption{Evolution of the specific angular momentum of galaxies in the three studied simulations and comparisons with specific angular momenta of galaxies of different morphological type from \protect\citet{fall_romanowsky13}. The panels show stars  (left), cold $T<10^4$ K gas (middle), and the total specific angular momentum of stars and cold gas (right). }
\label{fig:jms}
\end{center}
\end{figure*}

\subsection{Surface density profiles}
\label{sect:results_size}
Figure \ref{fig:ssgpro} shows the stellar and neutral gas surface density profiles for the simulated galaxies in the fiducial and \fesn simulations -- the two simulations that were evolved to $z=0$ --  along with the profiles of late type galaxies from the THINGS sample \citep{leroy_etal08} in the stellar mass range $M_\star\approx 3-10\times 10^{10}\ \rm M_{\odot}$ and the Milky Way \citep{McMillan2011,Dame1993}.  In the observed galaxies the gas surface density includes atomic and molecular gas corrected for Helium mass fractions, while in the simulations the gas mass includes all of the cold gas of $T<10^4$ K.

The figure shows that the fiducial simulation reproduces stellar surface density profile of  late type galaxies of this mass range, albeit with slightly larger bulge component than typical in the observed sample. Thus, this simulation does not just result in a stellar distribution with the correct half-mass radius, but also naturally reproduces a radial profile compatible with observed stellar distributions. In contrast, the surface density profile in the \fesn simulation is very different from all of the observed THINGS galaxies in this stellar mass range;  the $\Sigma_\star(R)$ profile is close to exponential, but as discussed above the scale length is considerably larger than most disk galaxies. Thus, it is a counterpart to the rare low-surface brightness galaxies of which some feature very extended exponential disk \citep[e.g.,][]{mcgaugh_bothun94}. Nevertheless, the number density of such galaxies, is roughly two orders of magnitude smaller than the number density of typical $L_\star$ galaxies we are investigating. Such galaxies are thus unlikely to form in a single random realization of a MW-sized halo. 

The gas surface density profile of the fiducial simulation does not match the profile of the Milky Way. However, its shape matches the overall shape of the gas surface density profiles in this mass range.
The normalization is somewhat higher than most THINGS galaxies which indicates that the simulated galaxy may be more gas rich than typically observed, and we return to the implications of this in \S\,\ref{sect:discussion}. 

The $\Sigma_{\rm gas}(r)$ profile in the \fesn simulation has an even higher normalization. Overall, this galaxy has lower star formation rate than the fiducial simulation at low redshifts, $0.5\lesssim z\lesssim 2$ \citep{AgertzKravtsov2015} and it retains more gas. The shape of the gas surface density profile is somewhat steeper than the profiles of observed galaxies. Profiles in both simulations exhibit more fluctuations than the $\Sigma_\star(R)$ due to the presence of spiral arms and other inhomogeneities in the gas distribution. Although the distribution of young stars exhibits similar fluctuations, the stellar mass distribution is considerably smoother which is reflected in the smoother $\Sigma_\star(R)$ profile. Overall, 
radial fluctuations of $\Sigma_{\rm gas}(R)$  in simulated galaxies are comparable to those in the data, which indicates a similar level of structure.

\subsection{Evolution and scaling of angular momentum}
\label{sect:results_am}

Figure \ref{fig:jms} shows the evolution of stellar, cold gas ($T<10^4\K$) and joint specific angular momentum\footnote{The specific angular momentum is computed as $j=\sum_i m_i\,\vec{r}_i\times\vec{v}_i/\sum m_i$ where summation is performed over all $i$ cells of cold gas or all star particles, and $\vec{r}$ and $\vec{v}$ are position and velocity relative to the galaxy's center of mass.}, of the central galaxies in the three simulations as a function of stellar mass. The models are compared to specific angular momenta of galaxies of different morphological types from \cite{fall_romanowsky13}, where the authors estimate the disk-only contribution via $j=2V_{\rm rot}R_{\rm d}$, where $R_{\rm d}$ is the scale radius of the exponential disc and $V_{\rm rot}$ is the asymptotic rotation velocity observed at large radii. 

The $M_{200}-M_\star$ and $M_\star-r_{1/2}$ relations in Figure~\ref{fig:lfms} and \ref{fig:rhms} indicated a monotonic relation between feedback strength and final galaxy size; as the strength of feedback increases, $r_{1/2}$ increases and $M_\star$ is decreased at any given value of $M_{200}$. We find that the same simple behavior is not recovered for the stellar mass - specific angular momentum relation ($M_\star-j$), as clearly visible for all ISM components in Figure\,\ref{fig:jms}. 

The galaxy in the fiducial run has a specific stellar angular momentum $j_\star\lesssim 100 \kmsec\kpc$  during early stages when $M_\star\lesssim 10^{10}\Msun$ ($z\gtrsim 1.5$). Angular momentum before that epoch fluctuates due to repeated mergers with the main progenitor leading to bursts of star formation and vigorous outflows \citep[see][for a thorough discussion]{AgertzKravtsov2015}. The last major merger occurs around $z\sim 1.8$, after which an epoch of disk formation begins, correlating with an increase in $j_\star$ by almost an order of magnitude. The corresponding increase in $r_{1/2}$ is only a factor of $\sim 2$ as discussed above. 
The final specific angular momentum of the galaxy in the fiducial run is compatible with spiral galaxies of roughly Sa-Sab type. 

In the run with inefficient feedback, $j_\star$ is always $\lesssim 100 \kmsec\kpc$, in agreement with early results on the ``angular momentum problem" \citep[e.g.][]{navarrosteinmetz00}. This does not come as a surprise, as the specific angular momentum content of accreting gas is expected to be smaller at higher redshifts, for a given galaxy progenitor \citep[e.g.][]{Pichon2011,Danovich2014}, and this galaxy formation model overpredicts the fraction of stellar mass forming out of this gas at early times instead of ejecting it via galactic winds \citep[see also][]{Ubler2014}. In addition,  the fraction of mass in the central galaxy is above the stability threshold for disks against an $m=2$ instability and bar formation. Such instabilities channel the gas towards the center, where it forms stars leading to a lower $r_{1/2}$. We discuss this issue in more detail in Section \ref{sec:stability} below. 

Rather surprisingly, the boosted feedback model fails to produce a galaxy rich in angular momentum, with $j_\star\sim 100-500 \kmsec\kpc$,  with large scatter, \emph{despite} the large physical extent as characterized by $r_{1/2}$. In contrast to the effect of feedback on the $M_\star-r_{1/2}$ evolution, the interaction between feedback and specific stellar angular momentum is more complex. In the case of 
the  {\tt ALL\_Efb\_e001\_5ESN} run, strong feedback results in a more spheroidal distribution of stars with angular momenta of individual orbits partially canceling each other. 

These results imply that  {\it strong feedback is not a panacea against the angular momentum catastrophe}, as different models of the star formation -- feedback cycle can give almost identical results in terms of global quantities ($M_\star$, SFR, $v_{\rm circ}$), but differ significantly in other, more detailed, observables \cite[][]{AgertzKravtsov2015}. In \S\,\ref{sect:recentstudies} we discuss feedback implementations further, and compare our results to recent studies on the angular momentum content of simulated galaxies.

We find that the angular momentum in the cold gas component ($j_{\rm gas}$) is always greater than the stellar counterpart. This is not surprising as stars form in the densest inner regions of the gaseous disk. Without stellar feedback, $j_{\rm gas}$ reaches $\sim 1000\kmsec\kpc$ at $z\sim 1.5$, 10 times higher than the corresponding $j_\star$. In the boosted feedback case, $j_{\rm gas}\sim 500\kmsec\kpc$ after the last major merger (i.e. for $M_\star>10^{10}\Msun$), with significant scatter. The fact that $j_{\rm gas}$ here is smaller than that in the fiducial model, by a factor of a few, may stem from how stellar feedback driven outflows, in this particular model, not preferentially ejects angular momentum poor gas \citep[as demonstrated by][]{Brook2011,Ubler2014}, but also high angular momentum  gas, hence destroying the cold ISM of the galaxy; by artificially boosting the available energy per SN, even gas associated with regions of low $\Sigma_{\rm SFR}$, which otherwise would not necessarily be as affected by feedback, can be ejected from the galaxy.  Overall, in the runs with feedback $j_{\rm gas}$ evolves approximately along the $j-M_*$ relation of observed galaxies. 

In the fiducial model, $j_{\rm gas}\sim 2000\kmsec\kpc$, a value compatible with observed $j_\star$ in late type spirals of Sc-Sd type. We note that the precise value for the simulated stellar angular momentum is sensitive to numerically resolving star formation at large radii. Furthermore, the cold gas extends much farther than the optical radius (see Figure \ref{fig:ssgpro}), in agreement with observations \citep[e.g.,][]{Kravtsov2013}.

\begin{figure}[t]
\begin{center}
\includegraphics[width=0.45\textwidth]{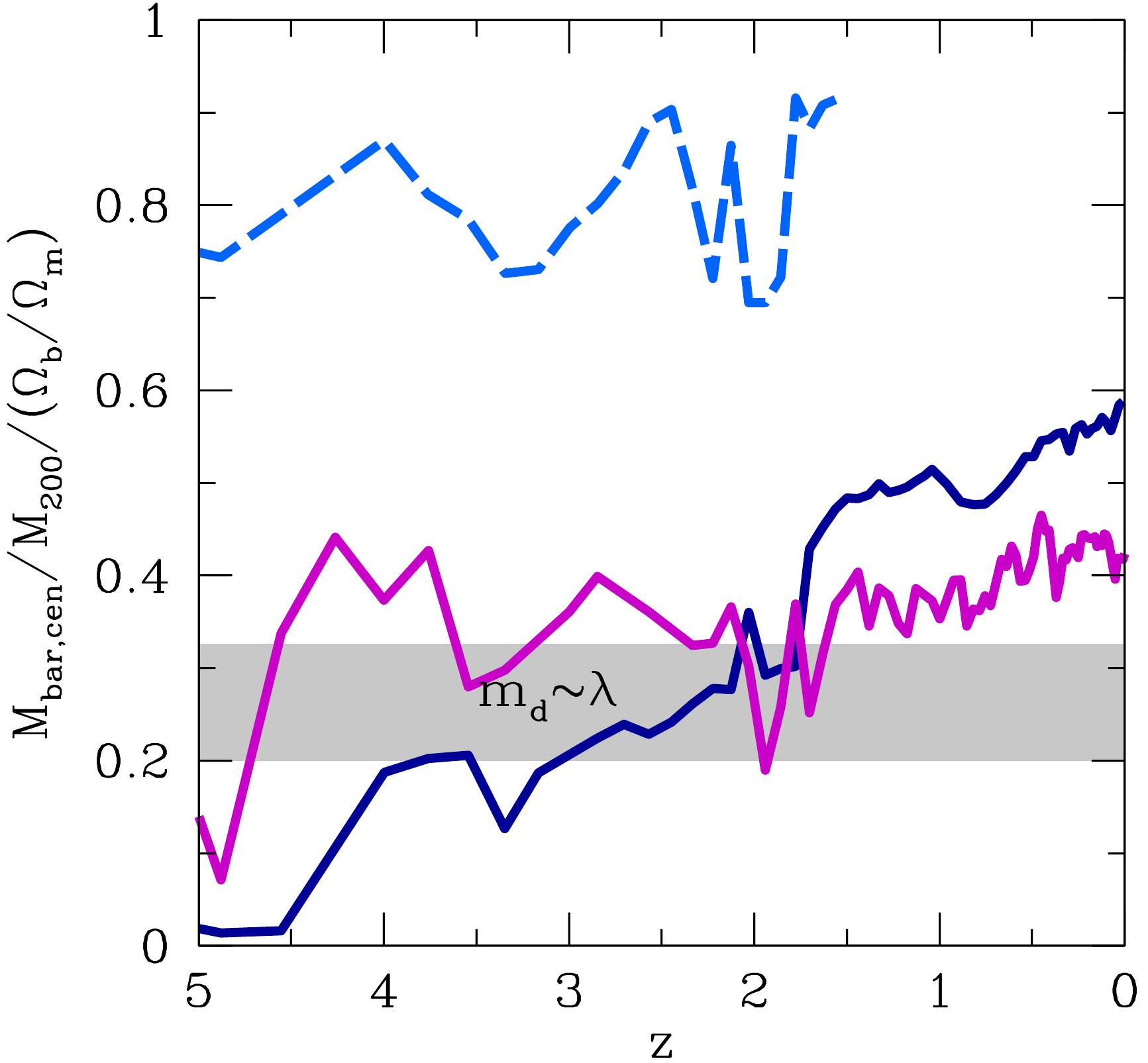}
\caption{The evolution of the halo baryon mass fraction (stars and cold gas) in the central galaxy in the three simulations (the colors and types are the same as in Figure~\ref{fig:jms}). The shaded region indicates a transition where the disk is expected to be susceptible to bar formation, leading to mass transport to the center (see text for detailed discussion).}
\label{fig:fbarz}
\end{center}
\end{figure}

\begin{figure*}[t]
\begin{center}
\includegraphics[width=\textwidth]{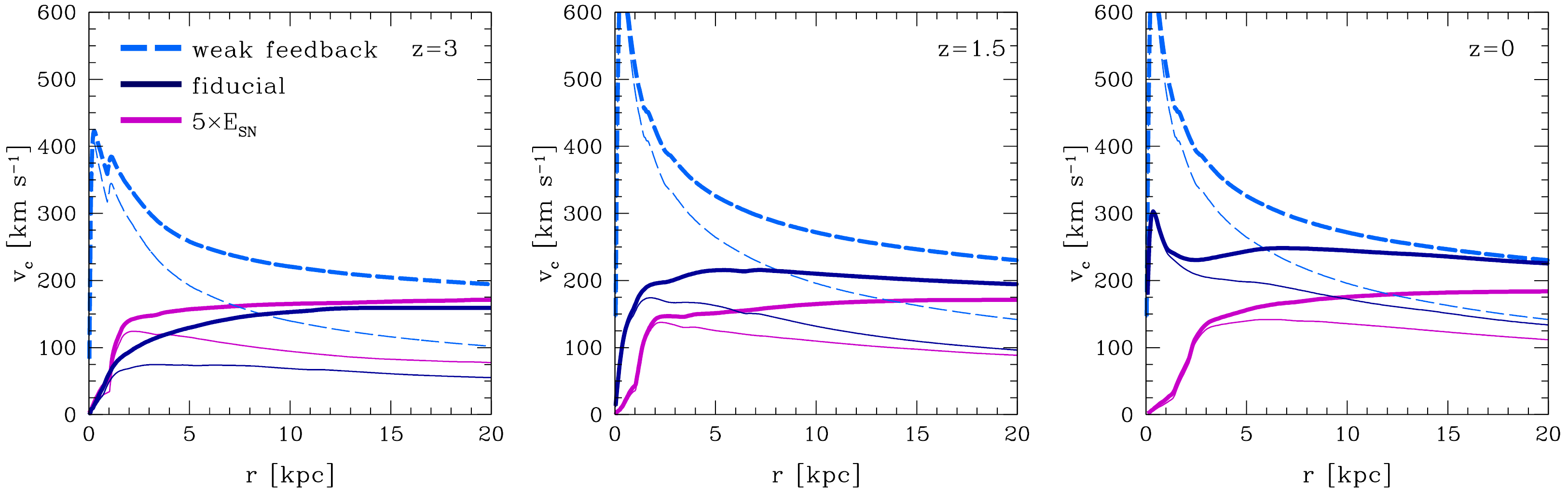}
\caption{Circular velocities ($v_{\rm c}=\sqrt{GM/r}$) for the three galaxy models at $z=3$ (left), 1.5 (middle) and 0 (right). The thick lines show $v_{\rm c}$ derived for all mass, whereas the thin lines account only for the stars and cold gas situated in the disk.}
\label{fig:vcirc}
\end{center}
\end{figure*}

\section{Discussion}
\label{sect:discussion}

The results presented in the previous section clearly show that the morphology of a galaxy forming in a given dark matter halo sensitively depends on the parameters of star formation and feedback. This conclusion is in agreement with a number of other recent studies, as we discuss below in Section \ref{sect:recentstudies}.

Although the role of feedback in setting galaxy morphology and basic properties, such as stellar mass and size, is widely acknowledged, the exact mechanisms by which feedback affects galaxy properties are not yet fully understood. Our results can be used to clarify some aspects of this question, as we discuss next.

\subsection{Disk instabilities, bulge formation and stellar feedback}
\label{sec:stability}
It is widely accepted that stellar and AGN feedback play key role in removing low angular momentum gas that is expected in the central regions of galaxies, even if there is no re-distribution of angular momentum during galaxy formation \citep[e.g.,][]{Bullock2001,vandenbosch2001,vandenbosch_etal01}. However, the role of feedback is likely to be more multi-faceted. Even if cosmologically accreted gas has initially no low angular momentum component, redistribution of angular momentum due to disk instabilities can quickly create such component \citep[e.g.,][]{KormendyKennicutt2004,Bournaud07,Elmegreen08,WeinbergKatz2007}. If star formation proceeds in this low angular momentum gas, this results in a more massive central bulge and possibly a spheroid-dominated galaxy. If, on the other hand, stellar feedback removes most of the gas before it is turned into stars, the remaining disk may have a much smaller stellar mass much than in observed galaxies. The role of feedback should thus be to remove low angular momentum gas efficiently but still reproduce observed disk masses. As we argue below, such self-regulating behavior can probably be understood by considering overall disk stability. 

Disk stability is affected by a number of factors \citep[e.g.][]{jogsolomon84,romeo92,Christodoulou1995,rafikov01,Romeo2010,Agertz2015b}, but a key factor controlling the overall global stability is the the fraction of potential contributed by the disk relative to the spheroidal component \citep{Efstathiou1982}. This dependence was realized  by \citet{ostriker_peebles73}, who used it to argue that existence of dynamically cold galactic disks requires the existence of massive spheroidal halos around them. 

\cite{Efstathiou1982} used $N$-body simulations to show that for exponential discs embedded in spherical haloes the onset of the bar instability can be characterized by the criterion\footnote{It is worth keeping in mind that other factors such as the vertical stellar velocity dispersion and multiple disk components will affect this criterion, as pointed out by \citet{Athanassoula2008}. Thus, this criterion is approximate.} 
\begin{equation}
\label{eq:bar}
\epsilon_{\rm  m}\equiv\frac{V_{\rm max}}{\sqrt{GM_{\rm disk}/R_{\rm d}}}\lesssim 1.1,
\end{equation}
where $V_{\rm max}$ is the maximum rotational velocity, $M_{\rm disk}$ is the mass of the disk, and $R_{\rm d}$ the disk's exponential scale length. Due to its simplicity, this is a widely used criterion of instability in semi-analytical models of galaxy formation \citep[e.g.][]{MoMaoWhite98,DeLucia2011,Guo2011}. 

\cite{MoMaoWhite98} used a detailed disk formation model to show that the criterion in Equation \,\ref{eq:bar} 
implies disk instability when the disk baryon mass relative to the total halo mass, $m_{\rm d}\equiv M_{\rm d}/M_{200}$, is greater than the disk spin parameter $\lambda_{\rm d}$: $m_{\rm d}\gtrsim \lambda_{\rm d}$. If the disk has a specific angular momentum similar to that of the dark matter halo, its spin parameter is expected to have a log-normal distribution with the median of $\lambda\approx 0.04$ with a fractional scatter of $\sigma_{\ln\lambda}\approx 0.55$, regardless of halo mass and cosmic time \citep[e.g.,][]{Bullock2001,vitvitska_etal02,Bett2010}. This result implies that when galaxy growth increases the disk mass 
beyond this limit, strong disk instabilities will set in. 

The subsequent evolution will depend on the efficiency of feedback; if feedback removes the low angular momentum gas resulting from momentum re-distribution in the unstable disk, and lowers the disk mass below the critical limit, the galaxy may maintain a disk mass fraction close to the critical instability threshold. If feedback is inefficient, the galaxy will form a massive central spheroidal component.

These different scenarios are indeed realized in our runs with different star formation and feedback parameters. In Figure\,\ref{fig:fbarz} we plot the disk baryon fractions (stellar + cold gas) for the three simulated galaxies as a function of redshift, with the grey band roughly indicating a possible transition into a disk unstable regime, and hence bar formation, assuming $\lambda= j/\sqrt{2}R_{200}V_{\rm 200}=0.044$, which we measure for our halo at $z=0$. Note that the mass fraction includes all of the mass in the galaxy, while stability criterion should include only disk mass. We do not attempt to separate disk and spheroidal components and simply examine the total mass of the galaxy. The rationale is that if the entire mass of the galaxy was in a cold disk, the figure would shows us the regime in which the disk is expected to be unstable. 

The run with low efficiency of star formation does not have sufficiently strong feedback and produces a galaxy with $\sim 80\%$ of the cosmic baryon fraction in the disk at all times. The feedback in the \fesn run, however, does drive strong outflows at $z\gtrsim 1$ and maintains disk masses right around the critical instability threshold from $z \sim 5$ to $z=0$. The feedback in the fiducial run is even stronger at high $z$, which keeps the disk mass is well below the instability threshold until the last major merger at $z\sim1.8$. At lower $z$, outflows become less prominent \citep[][]{AgertzKravtsov2015,Muratov2015} and the disk mass fraction reaches $m_{\rm d}\equiv M_{\rm disk}/M_{200}\approx 0.07$ (or $\approx 0.4-0.5$ in units of $\Omega_{\rm b}/\Omega_{\rm m}$) at lower $z$. This could indicate that the fiducial model is susceptible to $m=2$ disk instabilities at late times, which are responsible for growth of its bulge.

Indeed, this is consistent with evolution of mass distribution traced by the circular velocity profile in Figure\,\ref{fig:vcirc}, which shows the circular velocity for all three runs at different $z$ for only the baryons in the disk and all matter separately. At $z=3$, \fesn and the fiducial runs both have disk masses below the critical instability threshold and correspondingly have circular velocity profiles without a central ``spike'', which is very pronounced in the profile for the weak feedback run. At $z=0$, on the other hand, the $v_{\rm c}(r)$ profile for the  fiducial model features a modest peak in $v_{\rm c}(r)$ at $r\lesssim 1 \kpc$ consistent with late evolution of this disk above the critical threshold, whereas the circular velocity in the \fesn run, in which the
galaxy mass never gets significantly larger than the threshold, maintains a slowly rising $v_{\rm c}(r)$ profile without a central concentration. This figure thus indicates that development of central concentrations in galaxies is related to the development of disk instabilities when the central disk mass exceeds the approximate threshold of Equation \ref{eq:bar} above. 

This sheds light on the long-standing issue of ``spiky'' circular velocity profiles in simulations \citep[e.g.,][and references therein]{mayer_etal08}: our results indicate that prevalent central spikes in $v_{\rm c}(r)$ profiles, even at the highest $z$, are due to the inability of a given feedback prescription of keeping the disk mass around or below the instability threshold. 

At the same time, our results show that preventing formation of the central concentration does not necessarily result in formation of a disk-dominated galaxy. The galaxy that forms in our \fesn run has maintained low disk mass and has exponential surface density profile, but has an irregular morphology and a disk size that is larger than expected for the galaxy of such stellar mass. This is because stellar feedback in this
run continues to strongly ``stir'' the gas in the disk and disrupt galaxy ISM with outflows and fountains down to the smallest redshifts. The ISM in this run is thus perpetually in a perturbed state, which prevents formation of a well-defined cold disk component \citep{Roskar2014}.  

The above analysis illustrates the dual effect of stellar feedback: it both removes the low-angular momentum gas and regulates disk stability. In late type $L\sim L_{\ast}$  galaxies,  i.e. galaxies with low bulge-to-disk-ratios or lack of bulges altogether, are predominately low mass galaxies \citep[$M_\star\lesssim 1-5\times 10^{10}\Msun$][]{NairAbraham2010}, feedback effects should  delicately balance between driving ISM turbulence and outflows and allowing thin disks to form. Such galaxies likely form in haloes of masses $\sim 10^{11}-10^{12}\Msun$ \citep[][]{Behroozi2013,kravtsov_etal14}. 

\cite{vandokkum2013} studied the structural evolution of MW-mass progenitors back to $z\sim 2.5$ by matching cumulative co-moving number densities in the 3D-HST and CANDELS Treasury surveys. The surface brightness profiles were found to be well represented by S\'{e}rsic profiles with average indices evolving from $n\sim 1$ (exponential) to $n\sim 3$ (more centrally concentrated) from $z\sim 2.5$ to $z\sim 0$, with $\sim 1$ dex associated increase in average stellar masses and a factor of $\sim 2$ increase in galaxy sizes. These observational trends are close to the evolution found in our fiducial galaxy simulation for $z\lesssim 2$, which develops a more centrally concentrated mass profiles at late times via global disk instabilities, with roughly a Sa-Sab $z=0$ Hubble type.

At lower stellar masses ($M_\star\lesssim 10^{10}\ \M_{\odot}$), the stellar fraction $M_\star/M_{200}$ decreases rapidly (see Figure \ref{fig:lfms}), which means that these galaxies are more stable against global disk instabilities which prevents formation of central mass concentration. Indeed, bulgeless disks are observed to be more abundant at these masses \citep[][]{NairAbraham2010}. However, they are also more susceptible to disturbances due to injection of momentum and energy by stellar feedback, which can result in irregular, rather than thin disk, morphology.  

\subsection{Comparison with previous studies}
\label{sect:recentstudies}

A number of recent studies have emphasized the outmost importance of stellar feedback in shaping basic properties and morphologies of galaxies. Following initial findings of \citet{NavarroWhite1994}, \citet{navarrosteinmetz00} and \citet{Abadi03b} have confirmed the fact that when gas is not removed by winds, or when galaxies are allowed to form stars at a high rate during early epochs of evolution, the resulting $z=0$ galaxies are too compact and have specific angular momentum that is too low compared to observations of late type galaxies. 

Studies by \citet{Okamoto05}, \citet{Governato07}, \citet{zavala_etal08} and \citet[][]{Scannapieco08} have confirmed  that the problem arises from the loss of angular momentum the cold, condensed baryons experience after they are accreted in clumps of stars in gas and that the loss of angular momentum by baryons is mitigated by increasing the efficacy of stellar feedback. 

Efficient feedback prevents baryons from condensing and accumulating in dense central ``blobs'' and delays star formation until lower redshifts, when the gas is able to settle into a gaseous disk without significant loss of angular momentum \citep{maller_dekel02}. In particular, 
\citet[][see also \citealt{sales_etal10}, \citealt{PedrosaTissera2014}, and \citealt{Genel2015}]{zavala_etal08} have pointed out that agreement with observations requires that baryons that end up in late-type galaxies have specific angular momentum comparable to that acquired via gravitational torques by their host dark matter halos, as envisioned in the classical analytic models of disk formation \citep[][although see \citealt{dutton_vandenbosch12} who concluded that observed angular momenta of disk galaxies are somewhat lower than expected for their DM halos]{FallEfstathiou80,dalcanton_etal97,MoMaoWhite98}. The same conclusion can be reached by comparing sizes of galaxies to the expectations of such analytic models \citep[see][]{Kravtsov2013}.

The importance of efficient feedback in setting galaxy angular momentum, size, and morphology was confirmed by a number of recent studies \citep{sales_etal10,Aquila,Brook2011,Aumer2013,aumer_etal14,Ubler2014,Christensen2014,Christensen2014b,Genel2015,murante_etal15}. Thus, the role of feedback and the importance of proper modelling of the star formation--feedback loop are now firmly established.  The exact mechanisms by which feedback affects galaxy properties are nevertheless still a subject of debate.

One possible effect of feedback is to self-regulate star formation, making it inefficient and maintaining galaxies gas rich. Disks in gas rich galaxies are much more resilient during mergers and a thin disk can reform even after major mergers \citep{robertson_etal06}. This effect is likely significant at high redshifts when the age of the universe is comparable or smaller than the typical gas consumption time scale in galaxies. However, by itself self-regulation cannot result in formation of realistic disks as the accumulating gas eventually turns into stars, and without gas removal via winds, galactic stellar masses will end up too high to be compatible with observations.

\citet{zavala_etal08} and \citet{Scannapieco08} argued that the role of feedback was primarily to prevent gas condensation into dense clumps in low-mass dark matter halos and delaying gas accretion onto disk until later epochs, as argued by \citet{maller_dekel02}. \citet{Brook2011} and, more recently, \cite{Ubler2014} additionally argued that stellar feedback can promote disk formation by preferentially ejecting low angular momentum gas at early times ($z\gtrsim 1$), which can accumulate due to disk instabilities or angular momentum loss in mergers. \citet{Genel2015} also emphasized 
the importance of stellar feedback driven winds with large loading factors for the formation of angular momentum rich disks. They concluded, however, that AGN feedback via radio bubbles in their implementation leads to a net loss of angular momentum, concluding that AGN feedback may help explain low angular momentum of spheroidal stellar systems. Overall, \citet{Genel2015} concluded, as did another recent simulation studies by \citet{Aumer2013} and \citet{Fiacconi2015}, that diverse galaxy morphologies observed in their Illustris simulations did depend on the merger history of galaxies \citep[as argued since the earliest galaxy formation simulations][]{NavarroBenz91}. Both disk and spheroidal galaxies in their simulations had similar specific angular momentum at high $z$  comparable to that of their dark matter halos. Spheroidal galaxies, however, lost significant fraction of their angular momentum in major mergers. 
\cite{Sales2012}, however,  argued that alignment of angular momentum of accreting gas was a more important factor in setting galaxy morphologies than the merger history. In reality, it is likely that both mergers and angular momentum alignment of accreting gas affect galaxy morphology and structure \citep[see, e.g.,][]{aumer_etal14}.  

Our conclusions on the high importance of stellar feedback in setting properties of galaxies are in broad agreement with results of the studies discussed above. Our study have
contributed to our continuing quest to understand effects of feedback in several ways. First, our simulations implement a number of state-of-the-art prescriptions for star formation
and stellar feedback, such as molecular hydrogen based star formation and and implementation of both ``early'' (pre-supernova) and supernova momentum and energy injections. Most importantly, our simulations reach a resolution of $\sim 50-100$ pc in the ISM, allowing for the implementation of star formation and feedback prescriptions to operate on these scales, hence more closely resolving  the internal processes that shape galaxies structurally. 

Large-scale galactic winds then emerge (if they do) self-consistently due to coherent effects of local feedback from multiple star formation sites. This is in contrast to many recent studies, in which large-scale winds are launched explicitly as part of the overall feedback prescription, often accompanied by turning off the interactions between the wind and the galaxy's ISM and circumgalactic gas. For example, at $z = 0$, gravitational forces in the Illustris simulation analyzed by \citet{Genel2015} are softened on scales of $\sim 710\pc$, compared to $\sim75\pc$ in the our current work. Similar difference in resolution exists with the EAGLE and the Magneticum Pathfinder simulations \citep{schaye_etal15,Teklu2015}. The flip side, however, is that the computational expense demanded by such high resolution allowed us to 
simulate only a handful of galaxy models. 

Our results indicate that in addition to the role of feedback in ejecting low angular momentum gas pointed out in previous studies, feedback may also play a role in limiting disk mass, thereby suppressing disk instabilities and preventing associated loss of angular momentum in the first place. 

At the same time, our results show that efficient feedback that suppresses galaxy mass to the observed levels or below is no guarantee of forming a galaxy with realistic properties. Thus, feedback in our {\fesn} run maintains a low star formation rate and stellar mass, but the galaxy that forms is highly irregular and has size that is too large compared to observed galaxies. This is reminiscent of the recent results of \citet{Roskar2014}, who demonstrated the excessively violent energy and momentum input can lead to overheating of gaseous disks, even as the overcooling problem is alleviated \citep[see also][]{Stinson2013b}. This overheating then produces disk galaxies that are much thicker than observed. Although gas stirring by feedback may actually play a role in setting the disk thickness \citep{bird_etal13}, these results illustrate that stirring effects should be limited as not too thicken the disk excessively. 
 
Overall, our results show that basic properties of galaxy forming in a given halo are highly sensitive to the implementation and parameters of star formation and stellar feedback, consistent
with findings of \citet{Okamoto05} and \citet{Aquila}. 
This inherent sensitivity to details of the star formation--feedback loop  indicates that specific results, for a particular set of prescriptions, should be viewed with healthy skepticism, regardless of how well they reproduce particular set of observations. 

Our results published in a companion paper \citep{Liang2015}, in which we compared properties of circumgalactic gas in our simulations to observations, also highlight this point.
Although our fiducial run does result in many realistic galaxy features, including stellar mass, morphology, disk-to-bulge ratio, rotation curve shape, disk thickness \citep[see sections above and results in][]{AgertzKravtsov2015}, it fails spectacularly in reproducing observed properties of circumgalactic gas around galaxies of similar mass. In particular, its gaseous halo is too hot and is almost devoid of any extended warm gas component, the existence of which is revealed by ubiquitous absorption lines of low-ionization energy ions, such as Mg\,II. 

The model with boosted energy per supernova ({\fesn}) produces a circumgalactic medium closer to observations due to its continuing large-scale outflows to low $z$, fails to produce a realistic galaxy.  Although the discrepancy with observed CGM properties may be due, in part, to low spatial numerical in galactic halo regions \citep{Muzahid2014,Crighton2015} or missing physical processes such as cosmic rays \citep{Ensslin2007,Booth2013,SalemBryan2014}, the fact that a comparison with the CGM properties 
provides ``orthogonal'' constraints  illustrates the importance in comparing as wide a range of galaxy properties as possible \citep[see also][]{Brook2012}, in particular with the properties of gaseous component both within galaxies and in their surrounding halos. 

Finally, our results show that changes of parameters related to the star formation and stellar feedback prescriptions can result in variations of galaxy size, $r_{1/2}$, by more than an order of magnitude, while the final specific angular momentum in all baryons ($j_{\rm star+gas}$, Figure\,\ref{fig:jms}) varies only by a factor of two. This shows that a comparison to \emph{both} galaxy sizes and angular momenta is warranted, as they provide complementary information. Size evolution as a function of galactic mass, in particular, will likely to be a useful and stringent constraint on galaxy formation models \citep[][]{vandokkum2013,aumer_etal14}.

\section{Conclusions}
\label{sect:conclusions}
In this study we use cosmological zoom-in simulations of galaxy formation in a MW-sized halo to investigate the evolution of galaxy sizes, morphologies and angular momenta in runs with different parameters of the star formation -- feedback loop implementation.  In particular, we focus on three runs selected from a larger set of seven runs reported in \citet{AgertzKravtsov2015},  which employ the model for star formation and feedback described in detail in \citet{Agertz2013}. In these three runs the star formation and feedback prescriptions are kept the same, but the local star formation efficiency per free fall-time is varied by a factor of ten, from $\varepsilon_{\rm ff}=0.1$ in the fiducial run to $\varepsilon_{\rm ff}=0.01$ in the two low star formation efficiency runs. One
of the $\varepsilon_{\rm ff}=0.01$ runs has otherwise identical parameters to the fiducial run (inefficient feedback), while in the other the energy released by supernova is increased by a factor of five relative to the fiducial value of $10^{51}$ ergs.  Our main results and conclusions can be summarized as follows.

\begin{itemize}

\item[1.] One of the runs -- the fiducial run with the star formation efficiency of $\varepsilon_{\rm ff}=0.1$ -- generates vigorous large-scale feedback-driven outflows during early stages of galaxy evolution and reproduces the basic properties of late-type galaxies: morphology dominated by a kinematically cold disk, size of the stellar distribution, and specific angular momentum. In addition, the stellar and cold gas surface density profiles in this model is in good agreement with profiles of late type galaxies from the THINGS sample \citep[][]{leroy_etal08}.

\item[2.] The stellar half-mass radius in the fiducial simulation evolves approximately along the observed $z=0$ $r_{1/2}-M_{\ast}$ relation, and agrees well with the $r_{1/2}-M_{\ast}$ evolution for MW-like progenitors traced back to $z\sim 2.5$ by \cite{vandokkum2013} and \cite{Patel2013}. It also roughly follows the approximately linear $r_{1/2}-R_{200c}$ relation \citep[][]{Kravtsov2013}, especially if we correct 
for effects of pseudo-evolution on the halo mass and radius. The evolution along the observed $r_{1/2}-R_{200c}$ relation indicates 
that the specific angular momentum of the galaxy in this simulation is roughly comparable, or somewhat smaller, than the angular
momentum of the dark matter halo throughout galaxy evolution.

\item[3.] Our results indicate that in the fiducial model, feedback-driven outflows  delay the bulk of star formation to $z\lesssim 2$. After the last major merger, the galaxy enters an epoch of disk formation, with significant growth due to accretion from the hot gaseous halo. Here the specific angular momentum of the stars ($j_\star$) increases by an order of magnitude towards values compatible for galaxies of roughly Sa-Sab Hubble type.

\item[4.] Although the implementation of star formation and feedback in the run with inefficient local star formation ($\varepsilon_{\rm ff}=0.01$) is identical, large-scale outflows are absent, indicating that energy and momentum injection is too weak. The resulting galaxy suffers from the well-known overcooling and angular momentum problem, and does not resemble a realistic galaxy. We thus confirm conclusions of many previous studies that efficient feedback is critical to produce realistic late-type galaxies. 

\item[5.] Our results indicate that one of the roles of stellar feedback is to maintain stellar mass--halo mass fractions near or below the
critical threshold for global disk instabilities. We argue (see Section \ref{sec:stability}) that the failure of many previous simulations
in predicting the existence of late-type galaxies, with moderate or low bulge-to-disk ratios, in large part is related to the failure of maintaining the mass of the gaseous disk below such threshold, 
which leads to strong bars and spiral instabilities that efficiently and quickly channel gas towards the central regions. 

\item[6.] At the same time, we show that feedback can be too efficient and prevent the formation of a dominant, kinematically cold disk component. The run with $\varepsilon_{\rm ff}=0.01$, but boosted energy per supernova, produces a diffuse irregular galaxy with size much larger than a typical galaxy of that stellar mass, but with a rather low angular momentum. The galaxy in this run deviates from the observed $M_\star-r_{1/2}$ relation, which illustrates that increasingly abundant observations of the evolution of galaxy sizes will be a useful and stringent constraint on galaxy formation physics \citep[see also][]{Brooks2011,aumer_etal14}.

\item[7.] Although the galaxy produced in the fiducial run is far more realistic than the galaxy produced in the run with  $\varepsilon_{\rm ff}=0.01$, but boosted energy per supernova, our recent comparison of properties of circumgalactic gas
in these simulations to observations presented in \citet{Liang2015} shows that the fiducial run fails to reproduce the observed
CGM properties, while the run with boosted $E_{\rm SN}$, leading to an unrealistic galaxy, is a much close match to the observations. This illustrates the importance 
of testing star formation and feedback prescriptions on the full array of available data. 

\end{itemize}

Overall, the fact that the three runs in our study, differing by only a modest variation of star formation physics parameters,  produce galaxies of very different size, angular momentum and morphology within the same dark matter halo demonstrates the high sensitivity of the resulting galaxy properties to realizations of the star formation--feedback interplay \citep[see also][]{Okamoto05,Aquila}. 

Although this  sensitivity may appear to make the challenge of understanding galaxy formation daunting, we believe there is a reasonable hope of success. First, galaxy properties in high resolution simulations with efficient stellar feedback tend to self-regulate at the values close to observed properties of galaxies \citep[e.g.,][]{Hopkins2011,Hopkins2014}. Our results indicate this indirectly, as a galaxy with many realistic properties has emerged after just a few trials of two key parameters (i.e., the fiducial run). If the sensitivity to parameters was high and random, finding the region of parameter space that would produce a particular set of realistic properties would be quite difficult. Furthermore, from the differences between our fiducial and boosted SN energy runs we can conclude that energy and momentum injection has to be concentrated in regions of high star formation efficiency, not spread around throughout the disk \citep[see also][]{Governato2010}.  

At the same time, we see that some properties, such as the circumgalactic medium or disk thickness, are sensitive to details of the star formation--feedback loop implementation, and are likely even more demanding to capture in terms of  numerical resolution \citep[][]{House2011,Crighton2015}. Therefore, 
a careful evaluation and further development of such prescriptions warrants further attention and work.\\[5mm]

\acknowledgements
The simulations presented in this paper have been carried using the Midway cluster at the University of Chicago Research Computing Center, and we thank Douglas Rudd for his support. We thank Mike Fall for fruitful discussions and organizers of the KITP workshop ``Gravity's loyal opposition'' in May 2014 where some of the main results of this paper have been obtained. AK would like to thank the Simons foundation and organizers and participants of the Simons symposium on Galactic Super Winds in March, 2014, for stimulating and helpful discussions that aided in preparation of this paper. AK  was supported  by NASA ATP grant NNH12ZDA001N, NSF grant AST-1412107, and  by the Kavli Institute for Cosmological Physics at the University of Chicago through grants NSF PHY-0551142 and PHY-1125897 and an endowment from the Kavli Foundation and its founder Fred Kavli.

\bibliographystyle{apj}
\bibliography{simsize.bbl}

\end{document}